\documentclass[10pt,preprint]{aastex}
\newcommand{\vv}[1]{{\bf #1}}

\newcommand{\avg}[1]{{\langle{#1}\rangle}}
\newcommand{\Avg}[1]{{\left\langle{#1}\right\rangle}}
\def\simless{\mathbin{\lower 3pt\hbox
	{$\,\rlap{\raise 5pt\hbox{$\char'074$}}\mathchar"7218\,$}}} % < or of order
\def\simgreat{\mathbin{\lower 3pt\hbox
	{$\,\rlap{\raise 5pt\hbox{$\char'076$}}\mathchar"7218\,$}}} % > or of order

\newcounter{thefigs}
\newcommand{\fignum}{\arabic{thefigs}}

\newcounter{thetabs}
\newcommand{\tabnum}{\arabic{thetabs}}

\slugcomment{Submitted to \apj}

\shorttitle{Blanton (2000)}
\shortauthors{Segregation of Galaxy Types}

\begin{document}


\title{How Stochastic is the Relative Bias Between Galaxy Types?}


\author{Michael Blanton}
\affil{Princeton University Observatory, Princeton, NJ 08544-1001}
\affil{\and NASA/Fermilab Astrophysics Center\\
Fermi National Accelerator Laboratory, Batavia, IL 60510-0500}
\email{blanton@fnal.gov}


\begin{abstract}
Examining the nature of the relative clustering of different galaxy
types can help tell us how galaxies formed. To measure this relative
clustering, I perform a joint counts-in-cells analysis of galaxies of
different spectral types in the Las Campanas Redshift Survey (LCRS).
I develop a maximum-likelihood technique to fit for the relationship
between the density fields of early- and late-type galaxies. This
technique can directly measure nonlinearity and stochasticity in the
biasing relation.  At high significance, a small amount of
stochasticity is measured, corresponding to a correlation coefficient
$r\approx 0.87$ on scales corresponding to 15 $h^{-1}$ Mpc spheres. A
large proportion of this signal appears to derive from errors in the
selection function, and a more realistic estimate finds $r\approx
0.95$. These selection function errors probably account for the large
stochasticity measured by \citet{tegmark99a}, and may have affected
measurements of very large-scale structure in the LCRS. Analysis of
the data and of mock catalogs shows that the peculiar geometry,
variable flux limits, and central surface-brightness selection effects
of the LCRS do not seem to cause the effect.
%Although it is likely that selection effects which I cannot
%adequately model are actually causing the discrepancy, I cannot rule
%out that the properties of galaxies are genuinely different in the low
%redshift portion of the LCRS than in the rest of the survey.
\end{abstract}

%
% Introduction and motivation
%

\section{Motivation}
\label{c5_motiv}

Galaxies of different morphologies have different spatial
distributions, as first noted by \citet{hubble36a}. Early-type
galaxies, such as ellipticals and S0s, are highly clustered and
account for 90\% of galaxies in the cores of rich clusters; late-type
galaxies, such as spirals and irregulars, are less clustered and make
up 70\% of galaxies in the field (\citealt{dressler80a};
\citealt{postman84a}; \citealt{whitmore93a}). A general way of
expressing the relationship between the density fields of galaxies of
different types on any scale $R$ is with the joint probability
distribution $f(\delta_e, \delta_l)$; that is, the probability at any
location of finding an overdensity $\delta_e$ of early-type galaxies
and an overdensity $\delta_l$ of late-type galaxies. This quantity is
analogous to the joint probability distribution of galaxy and mass
density introduced by
\citet{dekel99a}.

The traditional method of measuring the properties of
$f(\delta_e,\delta_l)$ has been to compare the amplitude of the
fluctuations in each density field, using the correlation functions or
the power spectra. By these measures, the level of fluctuations in
ellipticals is stronger than that of spirals by a factor of 1.3--1.5
(\citealt{davis76a};
\citealt{giovanelli86a}; \citealt{santiago92a}; \citealt{loveday96a};
\citealt{hermit96a}; \citealt{guzzo97a}). These relative clustering 
properties are successfully reproduced by current models of galaxy
formation. For example, \citet{blanton99a} examined hydrodynamical
simulations and identified galaxies as dense, rapidly cooling clumps
of gas. Older galaxies, which correspond to early-types, turned out to
be clustered more strongly than younger galaxies, which correspond to
late-types. The relative bias factor, $b\equiv\sigma_e/\sigma_l$,
where $\sigma^2\equiv\avg{\delta^2}$, is approximately 1.5 between
these populations.  Semi-analytic models, which follow halos in
collisionless $N$-body simulations and use simple models for
star-formation and feedback inside each halo and for the effect of
halo mergers, find similar results (\citealt{somerville99a}).

However, \citet{blanton99a} also found that there was considerable
scatter between the two density fields; that is, that there was not a
one-to-one relationship between the number of old galaxies in a region
to the number of young galaxies. A measure of this scatter is the
correlation coefficient
$r\equiv\avg{\delta_e\delta_l}/\sigma_e\sigma_l$ between the
early-type overdensity field $\delta_e$ and the late-type overdensity
field $\delta_l$. In the simulations, $r \sim 0.5$--$0.8$.  On the
other hand, the semi-analytic models of \citet{somerville99a} find
that the correlation coefficient $r\sim 0.9$; that is, they find very
little scatter.  The essential difference between the predictions of
this model and that of the hydrodynamic model is the effect of the
temperature history of the gas in the hydrodynamic simulations and its
relationship with large-scale structure. Thus, one can use the
correlation coefficient between different galaxy types to distinguish
between these models of galaxy formation.

Measuring this scatter requires a probe of $f(\delta_e|\delta_l)$
which differs from the traditional statistics mentioned above.  For
example, two completely unrelated density fields ($r=0$) can have the
same correlation function. To detect the scatter, one must compare the
density fields point by point, not just compare the overall levels of
the fluctuations.  A direct approach to constraining the properties of
$f(\delta_e,\delta_l)$ is to measure the related joint probability
distribution $P(N_e, N_l)$ of finding $N_e$ early-type and $N_l$
late-type galaxies in a single cell of size $R$. After all, this
latter probability is simply $f(\delta_e,
\delta_l)$ convolved with Poisson distributions. If one notes that
\begin{equation}
\label{c5_bayes}
f(\delta_e,\delta_l) = f(\delta_l|\delta_e) f(\delta_e),
\end{equation}
then one can write
\begin{eqnarray}
\label{c5_prob2}
&P(N_e, N_l)& = \int d\delta_e 
\frac{{N}_{e,\mathrm{exp}}^{N_e} (1+\delta_e)^{N_e}}{N_e!} e^{-{
N_{e,\mathrm{exp}}}(1+\delta_e)} f(\delta_e) \cr
&&\times \int d\delta_l 
\frac{{N}_{l,\mathrm{exp}}^{N_l} (1+\delta_l)^{N_l}}{N_l!} e^{-{
N_{l,\mathrm{exp}}}(1+\delta_l)}
f(\delta_l| \delta_e),
\end{eqnarray}
where ${N}_{e,\mathrm{exp}}$ and ${N}_{l,\mathrm{exp}}$ are the
average number of galaxies of each type expected in a cell of a given
volume (and given selection criteria). 
%Figure \ref{c5_distrib2} shows
%an example of the $P(N_e, N_l)$ resulting from the convolution with
%Poisson noise of a simple form of $f(\delta_e, \delta_l)$, namely, a
%log-normal distribution of $\delta_e$ and deterministic linear bias
%between $\delta_e$ and $\delta_l$ (see Section \ref{c5_method} for
%details). 
Naturally, one can integrate Equation (\ref{c5_prob2}) over
$N_l$ to obtain:
\begin{equation}
\label{c5_prob1}
P(N_e) = \int d\delta_e 
\frac{{N}_{e,\mathrm{exp}}^{N_e} (1+\delta_e)^{N_e}}{N_e!} e^{-{
N_{e,\mathrm{exp}}}(1+\delta_e)} f(\delta_e),
\end{equation}
the probability distribution of counts of early-type galaxies. As I
show below, one can use Equation \ref{c5_prob2} to devise a maximum
likelihood method to fit for $f(\delta_l| \delta_e)$, and Equation
\ref{c5_prob1} to fit for $f(\delta_e)$.

Equation (\ref{c5_prob2}) provides a direct probe of the relationship
between galaxy density fields $f(\delta_l|\delta_e)$, including its
nonlinearity and scatter.  Consider for contrast the work of
\citet{benoist99a}, who infer nonlinearity in the relative bias of
galaxies of different luminosities in the Southern Sky Redshift Survey
from the scale dependence of the higher-order moments of the density
fields. Using the same data, one could instead compare the observed
$P(N_e, N_l)$ to models and detect nonlinearity more directly.
Furthermore, the joint distribution contains more information than a
comparison of the moments of each density field. While moments of the
density field yield averaged information about the fluctuations,
$P(N_e, N_l)$ yields a point-by-point comparison of two density
fields, which can be much more powerful (\citealt{santiago92a}). For
instance, one can use this comparison to determine whether the effects
detected by
\citet{benoist99a} are actually due to nonlinearity (as they propose),
or perhaps due properties of the scatter in the relationship between
low luminosity and high luminosity galaxies.

In this paper, I perform an maximum-likelihood analysis of this
joint distribution for different spectral types of galaxies in the Las
Campanas Redshift Survey (LCRS), using cells with volumes approximately
equal to that of cubes 25 $h^{-1}$ Mpc on a side. 
A similar analysis has been performed on the LCRS by Tegmark \&
Bromley (1999; hereafter TB99), to which I will compare my results
% \citet{tegmark99a}
throughout.  Essentially, their method calculates the second moments
of $f(\delta_e,\delta_l)$, namely $\sigma_l^2\equiv\avg{\delta_l^2}$,
the variance of the density field of late-type galaxies,
$\sigma_e^2\equiv\avg{\delta_e^2}$, the variance of early-type
galaxies, and $r\equiv\avg{\delta_e\delta_l}/\sigma_e\sigma_l$, the
correlation coefficient between the two fields, which is unity if the
fields are perfectly correlated and zero if the fields are completely
uncorrelated. As I will show below, calculating second moments is
probably not sufficient on the scales which TB99 probe ($\sim 5$--$10$
$h^{-1}$ Mpc), because it does not correctly account for the fact that
density fields cannot be negative. For this reason, the resulting $r$
may overestimate the degree of scatter in the relationship between the
two fields. On larger scales where $\sigma\ll 1$, these differences
would of course be much reduced. Furthermore, the moments method also
yields no information on how nonlinear the relationship between the
density fields of the two galaxy types is, which the maximum
likelihood method described here will. Finally, I have detected
important effects concerning the galaxy selection function which
affect the results of this analysis and have consequences for the
interpretation of TB99 and other measurement of large-scale structure
in the LCRS.

This paper is organized as follows.  In Section \ref{c5_data}, I
describe the details of the LCRS.  In Section \ref{c5_selfunc}, I
describe the method used to calculate the selection function for the
survey.  In Section \ref{c5_method}, I describe in detail the maximum
likelihood analysis of the counts-in-cells.  In Section
\ref{c5_results}, I present the results of fitting for the
relationship between the different galaxy types and demonstrate the
presence of systematic errors in the selection of galaxies in this
survey. In Section \ref{c5_mock}, I describe the results of an
analysis of mock catalogs, in order to quantify a number of possible
statistical and systematic effects as well as to evaluate the
importance of cosmic variance.  I conclude in Section
\ref{c5_conclusions}.

%
% Describe the LCRS data
%

\section{Galaxies in the Las Campanas Redshift Survey}
\label{c5_data}

The LCRS (\citealt{shectman96a}) consists of around 25,000 galaxy
redshifts with a median of $z\sim 0.1$, covering an area of about 700
deg$^2$ on the sky. Three long slices ($1.5^\circ\times80^\circ$) were
surveyed in the North Galactic Cap, and three in the South Galactic
cap. Within each hemisphere, the slices had the same right-ascension
limits but were separated by several degrees in the declination
direction. $R$-band photometry was obtained at the Las Campanas Swope
1m telescope using three different CCDs; spectra were taken at the Las
Campanas Du Pont 2.5m telescope, first with a 50-fiber MOS and later
with a 112-fiber MOS. The galaxies in the 50-fiber MOS fields were
selected between $16<m<17.3$; the galaxies in the 112-fiber MOS
fields were selected between $15<m<17.7$. In addition, a
magnitude-dependent central surface-brightness cut was applied in
order to avoid putting fibers onto galaxies unlikely to yield useful
spectra. This cut takes the form:
\begin{equation}
\label{c5_mcdef}
m_c < m_{c,\mathrm{cut}} - 0.5 (m_{\mathrm{max}} - m),
\end{equation}
where $m_{\mathrm{max}}$ is the faint magnitude limit, $m_c$ is a
central aperture magnitude consisting of the flux within a two pixel
radius of the center of the image, and $m_{c,\mathrm{cut}}$ is 18.85
for the 112-fiber fields and 18.15 for the 50-fiber fields.  Since
each of the CCDs had a different pixel size, the size of the aperture
with which $m_c$ is calculated varies within the survey between about
3$''$ and 4$''$ in diameter. The central magnitude cut excludes about
12\% of the detected galaxies.  As in all redshift surveys, this
surface brightness cut can affect the relationship between the
luminosity function and the selection function, since the selection is
not purely based on apparent magnitude. Below, I test the dependence
of my results on this cut.

\begin{figure}
\figurenum{\fignum a}
\plotone{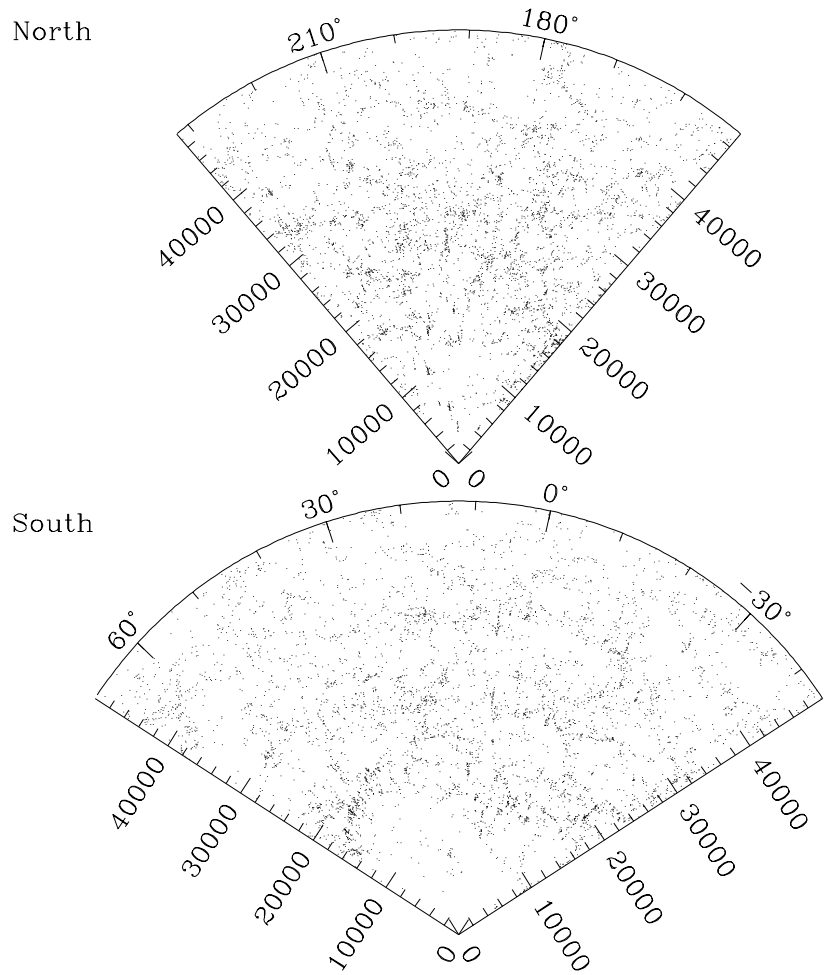}
\caption{\label{c5_lcrse} Distribution of early-type galaxies in the
LCRS. Radius indicates $cz$ in km/s; angle indicates right ascension.}
\end{figure}

\begin{figure}
\figurenum{\fignum b}
\plotone{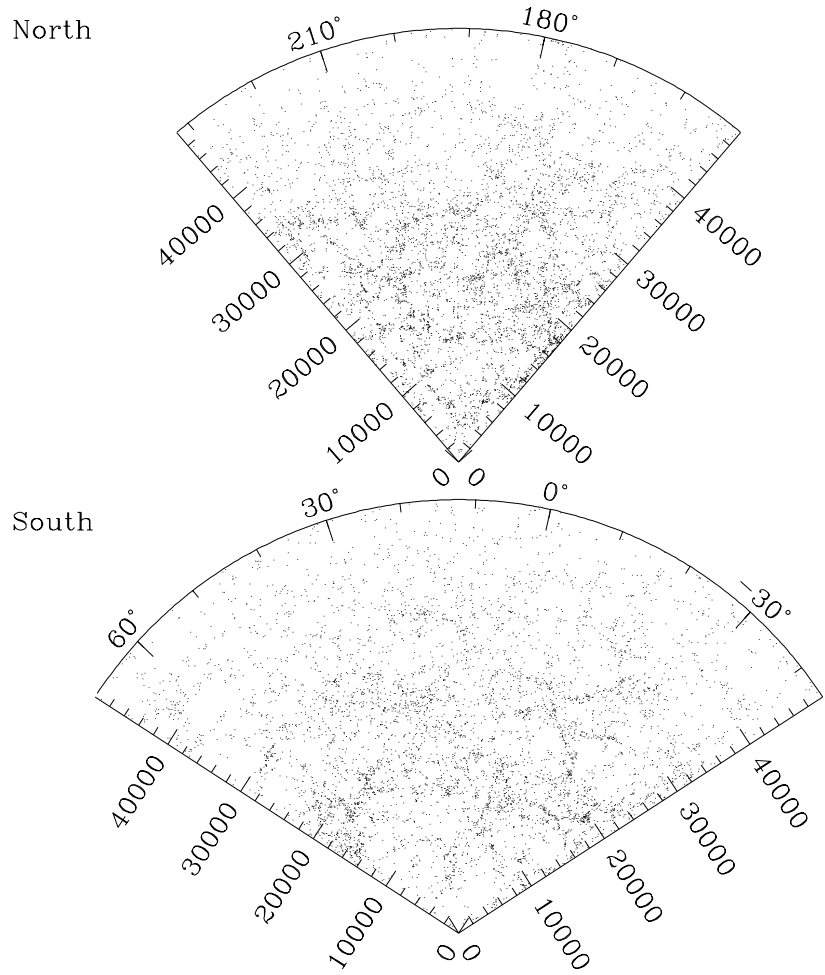}
\caption{\label{c5_lcrsl} Same as Figure \ref{c5_lcrse}, for the
late-type galaxies.}
\end{figure}

\citet{bromley98a} have used a spectral classification scheme to
divide the LCRS galaxies into six ``clans.''  Their method performs a
singular-value decomposition (SVD) on the set of galaxy spectra
(converted to rest wavelengths) to obtain an orthogonal set of galaxy
``eigenspectra.'' They found that the galaxy spectra form a
well-defined one-dimensional locus when projected onto the
two-dimensional plane defined by the two most significant
eigenspectra.  The ``clans'' are defined by each galaxy's position
along this locus.  The spectra of the ``late-type'' clans more closely
resemble the spectra of emission-line galaxies with young stellar
populations, while the spectra of ``early-type'' clans have more
prominent absorption features. Clans are labeled from 1 to 6, in order
of increasing ``lateness'' of the spectra. For my purposes, I will
split the galaxies into just two groups: an early-type group
consisting of clans 1 and 2 and a late-type group consisting of clans
3 through 6. I place absolute magnitude limits on the early-type group
of $-22.5<M<-18.8$ and on the late-type group of $-22.0<M<-18.5$.
Outside of these limits there are only a handful of galaxies and it is
risky to determine the luminosity function and to calculate the
selection function there.  This procedure yields about 10,000 galaxies
in each group.  I show the spatial distribution of each type of galaxy
in Figures \ref{c5_lcrse} and \ref{c5_lcrsl}.

\begin{figure}
\figurenum{\fignum}
\plotone{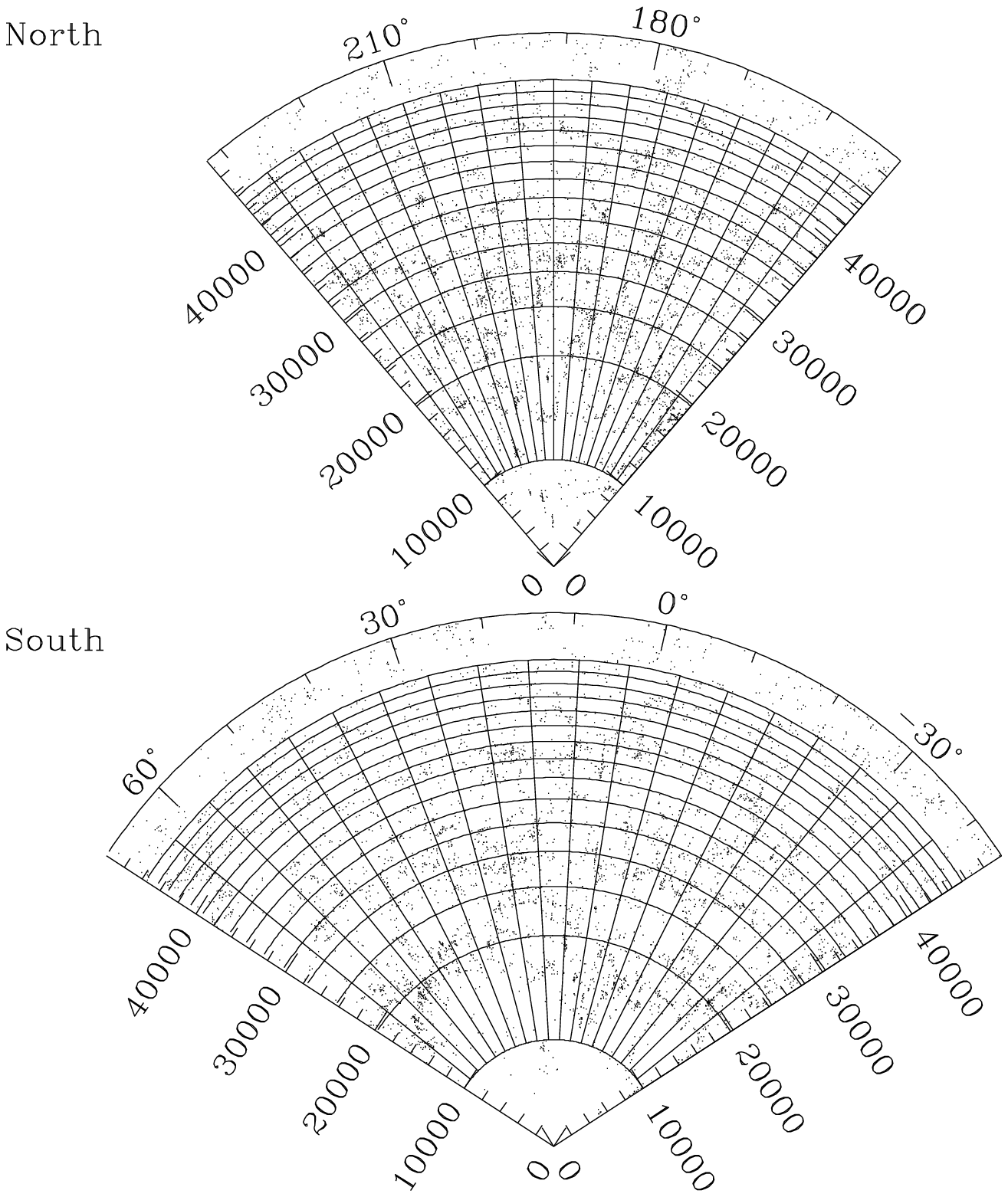}
\caption{\label{c5_arcs} Same as Figure \ref{c5_lcrse}, with
approximate cell boundaries superposed. Redshift shells are spaced
such that cells have constant volume. Angular boundaries shown are only
illustrative; the real angular boundaries are defined by the
configuration of the MOS fields (\citealt{shectman96a}) and are thus
somewhat more complicated.}
\end{figure}

The geometry of the LCRS complicates an attempt to perform a
counts-in-cells analysis on it.  To do so, I create 14 redshift shells, each
with an equal volume; thus, the shells at higher redshift have a
shorter radial extent. Figure \ref{c5_arcs} shows the boundaries of
these shells. In the angular dimension, I divide the survey into cells
which are 3 MOS fields on each side. In the right ascension direction,
the fields are adjacent; in the declination
direction, the fields from the three slices in each Galactic
hemisphere are combined.  The radial spokes in Figure \ref{c5_arcs}
are representative of the division in the right ascension direction,
although the actual cells are somewhat more complicated, because the
MOS fields are not all perfectly aligned in right ascension. This
procedure produces 518 cells total, each with a volume equivalent to a
15 $h^{-1}$ Mpc radius sphere, of about cubical dimensions at $z\sim 0.1$
(except for the gaps in the declination direction).  The average
number of galaxies in each cell of each type is about 20, meaning that
the average contribution of Poisson noise to the variance is about
$0.05$, though the actual contribution varies considerably with radius
due to the selection function. A reasonable variance of the underlying
density field at these scales in standard cosmological models is about $0.25$,
meaning the Poisson contribution to the variance is only about 20\%.

%
% Selection Function
%
\section{Selection Function}
\label{c5_selfunc}

The selection function for a flux-limited survey is the fraction of
galaxies in a given absolute magnitude range which are within the flux
limits at a given redshift. If one considers galaxies with
luminosities in the range between $L_{\mathrm{min},0}$ and
$L_{\mathrm{max},0}$, and take the luminosity function $\Phi(L)$ to be
normalized in that range, one can write the selection function:
\begin{equation}
\label{c5_sfeq}
\phi(z) = {\int_{L_{\mathrm{min}}(z)}^{L_{\mathrm{max}}(z)} dL\, \Phi(L)
f_g(m) f_{t}},
\end{equation}
where $L_{\mathrm{min}}(z)$ and $L_{\mathrm{max}}(z)$ are the minimum
and maximum luminosities visible at redshift $z$, given the flux
limits of the field under consideration, and $m$ is the apparent
magnitude corresponding to $L$ and $z$.  $f_t$ and $f_g(m)$ contain
information on the incompleteness of the survey. $f_{t}$ is the
sampled fraction of galaxies in the current field; that is, the
fraction of galaxies within the stated flux limits whose redshifts
were obtained, for each field. Galaxies are missed for a number of
reasons: the limited number of fibers, the central magnitude cut, the
fact that fibers cannot be placed closer than 55'', and the failure to
determine redshifts from spectra.  One must account for the fact that
these effects are not distributed evenly in apparent magnitude (in
practice, due mostly to the magnitude-dependence of redshift
failures). To do so, I adjust the probability of observing a galaxy by
a factor $f_g(m)$ which is the completeness fraction at each apparent
magnitude (normalized to unity, since $f_t$ already accounts for the
total number of missing galaxies).

Thus, one's task is to calculate the luminosity function given the
survey's various selection effects.  Although \citet{bromley98a} and
\citet{lin96a} have already published luminosity functions for
different galaxy types in the LCRS, for a number of reasons I was
motivated to reexamine the determination of the luminosity and
selection functions. In particular, some features of the selection
functions seemed suspicious; in fact, this suspicious behavior remains
in my analysis, and I will describe it in detail later.
I calculated the luminosity function using the standard iterative,
nonparametric maximum likelihood technique described in detail in
\citet{efstathiou88a}.  Their technique is based on maximizing the
probability $p(L|z)$ of having observed each galaxy of luminosity $L$
at its given redshift $z$. If the galaxy luminosity function is
$\Phi(L)$, a galaxy at redshift $z_j$ and having luminosity $L_j$ is
observed with a probability per unit redshift and luminosity of:
\begin{equation}
\label{c5_Lzjnt}
p(L_j, z_j) = \left\{
\begin{array}{cc}
\Phi(L_j) \rho(z_j) f_g(m_j) f_{t,j} & 
{\mathrm{if~}}m_{{\mathrm{min}},j}<m_j<m_{{\mathrm{max}},j}, \cr
0 & \mathrm{otherwise}.
\end{array}\right.
\end{equation}
Here, $f_g(m_j)$ and $f_{t,j}$ are as defined above, $\rho(z_j)$ is
the density of galaxies at redshift $z_j$, and $m_{{\mathrm{min}},j}$
and $m_{{\mathrm{max}},j}$ vary depending on what MOS field the galaxy
is in, since each field has different flux limits.  In order to relate
apparent magnitudes $m$ to absolute magnitudes $M$ and luminosities
$L$, I assume a flat universe with $\Omega_m=1$, which yields a
distance modulus:
\begin{equation}
\label{c5_dmeq}
DM(z) = m - M = 25 + 5 \log_{10}\left[x(1+z)\right]+K(z),
\end{equation}
where the comoving distance is
\begin{equation}
x = \frac{2c}{H_0} \left[1- \frac{1}{\sqrt{1+z}}\right],
\end{equation}
and the $K$-correction is of the form:
\begin{equation}
K(z) = 2.5\log_{10}(1+z).
\end{equation}
This $K$-correction is approximately equivalent to that which is
appropriate for Sbc galaxies (\citealt{frei94a}). Since the entire
analysis is independent of the value of $H_0$, I will for simplicity
assume the value of 100 km/s/Mpc.

Given the joint probability in Equation (\ref{c5_Lzjnt}), the
conditional probability of observing a galaxy of luminosity $L_j$
given its redshift $z_j$ is then:
\begin{equation}
\label{c5_Lzcond}
p(L_j| z_j) = \frac{p(L_j,z_j)}{p(z_j)} = \frac{\Phi(L_j) f_g(m_j)}
{\int_{L_{{\mathrm{min}},j}}^{L_{{\mathrm{max}},j}} dL \Phi(L) f_g(m)},
\end{equation}
where the limits on the integral are determined by the apparent
magnitude limits in each field, as implied by Equation
(\ref{c5_Lzjnt}). Note that this estimator is density-independent, as
all reference to a density field $\rho(z)$ or the angularly varying
sampling fraction $f_t$ drops out. This approach differs somewhat from
that of \citet{lin96a}, which weights the densest regions more
heavily. 

In the Appendix, I describe in detail the method for maximizing this
probability and for calculating the normalization of the luminosity
function $n_1$.  Using the derived luminosity function and its
normalization, one can calculate the selection function and, from
that, quantities such as the expected distribution of number with
redshift or the expected counts in cells. I perform this fit
separately for each combination of MOS field type and Galactic
hemisphere, and for the early- and late-type galaxies separately. I
will give results not in terms of $\Phi$ itself, but in terms of the
luminosity function expressed in logarithmic intervals of luminosity
and normalized to the average density:
\begin{equation}
\label{c5_hatphi}
\hat\Phi = n_1 \ln 10 L \Phi.
\end{equation}

%
% Describe the maximum likelihood method
%

\section{Fitting a Counts-in-Cells Distribution}
\label{c5_method}

Here I describe a maximum likelihood method for constraining the relative
clustering properties of early and late-type galaxies, assuming models
for the density distribution $f(\delta_e)$ and the relative bias
$f(\delta_l|\delta_e)$. I will assume $f(\delta_e)$ is distributed as
a log-normal. For $f(\delta_l|\delta_e)$, I describe a number of
deterministic models as well as a model which includes scatter.

\subsection{Defining the Likelihood}

Let us assume that one has divided a galaxy redshift survey into cells
and counted the numbers of early- and late-type galaxies, denoted
$N_{e,i}$ and $N_{l,i}$, in each cell $i$. (I describe in Section
\ref{c5_data} how I do so for the LCRS). Given the selection function
of the survey, which may depend on angle as well as redshift, I define
the expected count in cell $i$ of early-type galaxies:
\begin{equation}
{N}_{e,{\mathrm{exp}},i} = \int dV_i n_e \phi_e(r,\theta,\phi),
\end{equation}
where the integral is over the volume of cell $i$, and define $
N_{l,{\mathrm{exp}},i}$ similarly. Then, given a probability distribution
$f(\delta_e, \delta_l)$ characterized by a set of parameters
$\vv{\alpha}$, I define the likelihood for that cell as
\begin{equation}
\label{c5_like}
L_i \equiv P(N_{e,i}, N_{l,i} | \vv{\alpha}),
\end{equation}
where the quantitity on the right-hand side is determined by Equation
(\ref{c5_prob2}). I have implicitly assumed that all sets of
$\vv{\alpha}$ have equal Bayesian prior probability. I then minimize the
quantity
\begin{equation}
{\mathcal L} \equiv -2 \sum_i \ln L_i
\end{equation}
by varying the parameters $\vv{\alpha}$ in order to find the best fit.

In practice, it is time-consuming to fit simultaneously for the
parameters of both $f(\delta_e)$ and $f(\delta_l|\delta_e)$ in
Equation (\ref{c5_bayes}). Thus, I
first fit for the parameters of the density distribution
$f(\delta_e)$; I do so with an equation analogous to Equation
(\ref{c5_like}), using the probability $P(N_e|\alpha)$ given in
Equation (\ref{c5_prob1}) in place of $P(N_e, N_l|\alpha)$. Then I fix
the parameters of $f(\delta_e)$ and use Equation (\ref{c5_like}) to
fit separately for those of $f(\delta_l|\delta_e)$. 
Experiments show that the difference between this approach and
fitting for all the parameters simultaneously is negligible compared
to the error bars.

\begin{figure}
\figurenum{\fignum}
\plotone{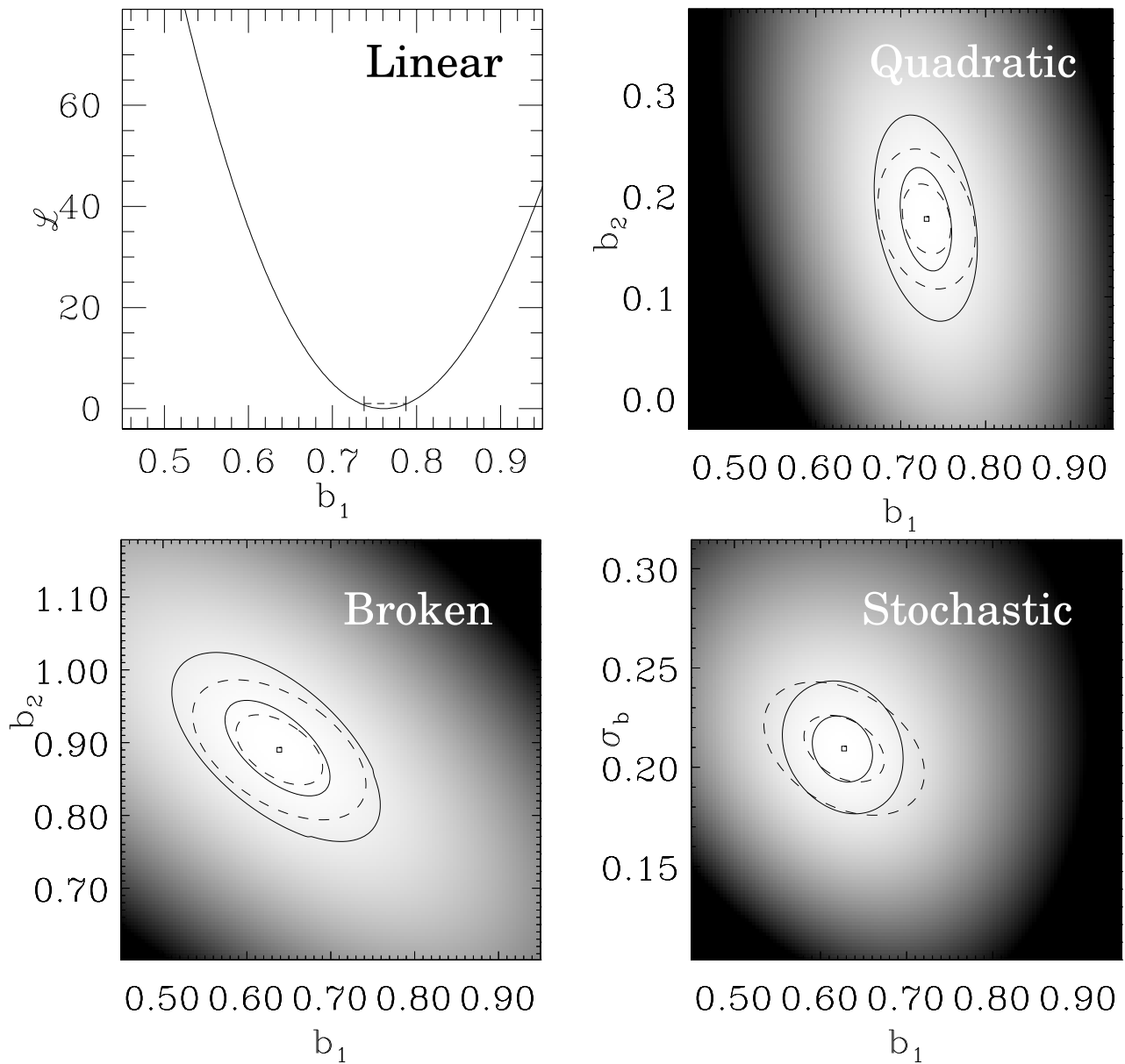}
\caption{\label{c5_biaslike} The likelihood functions of the linear,
quadratic, broken, and stochastic bias fits, as labeled, for each
galaxy type. In the top left panel, the likelihood is plotted against
$\sigma$; in the other panels, the likelihood is plotted as a
greyscale against $b_1$ and $b_2$ or $\sigma_b$. 
Shown as the dashed line in the
top left panel is the error bar as estimated by bootstrap; note that
these errors correspond closely to the errors one would estimate if
one defined the errors according to where ${\mathcal L} = {\mathcal
L}_{\mathrm{min}} + 1$. Similarly, in the other panels, the solid
lines show where ${\mathcal L} = {\mathcal L}_{\mathrm{min}} + 1$ and
${\mathcal L} = {\mathcal L}_{\mathrm{min}} + 4$, while the dashed
lines show the $1\sigma$ and $2\sigma$ error bars estimated by the
bootstrap method. Again, these two error estimates are quite
similar. }
\end{figure}

Once one has found the maximum likelihood fit to the parameters, one
would like to calculate the errors associated with them. I use three
methods of calculating error bars. First, I perform Monte Carlo
bootstrap estimates of the errors. Second, I again make a Monte Carlo
estimate, only now creating realizations based on the model to which I
have fit. (This procedure also serves to show that the method is
unbiased). Third, I simply look at the likelihood contour ${\mathcal
L}_{\mathrm{min}} + 1$ in order to estimate the 1$\sigma$ error
contour, which it is in the limit that ${\mathcal L}$ is a
paraboloid. All of these methods agree within 10--20\%, and the listed
error bars in this paper are those determined by bootstrap. (For a
comparison of the likelihood method and the bootstrap method, look at
%Figures \ref{c5_pdflike} and 
Figure \ref{c5_biaslike} in Section
\ref{c5_results}).

I would also like to compare different models using likelihood ratio
tests. For instance, one may want to fit a model with parameters
$(\alpha_1, \alpha_2)$ and ask whether this model is significantly
better than fitting a model with the single parameter $(\alpha_1)$. In
this case, one calculates the ``likelihood ratio'' between the models,
\begin{equation}
{l} = {\mathcal L}_{\mathrm{min}}(\alpha_1) -
{\mathcal L}_{\mathrm{min}}(\alpha_1, \alpha_2).
\end{equation}
To evaluate the significance of this likelihood ratio, I create a large
number of Monte Carlo realizations based on the single-parameter model
and ask how often one sees a likelihood ratio as large as the measured
$l$ purely by accident. I will make extensive use of this technique
below.

A general concern about using a counts-in-cells analysis in a
flux-limited survey ({\it e.g.}, TB99, \citealt{efstathiou90a}) is that
the measured density distribution in the cells may be affected by the
variation of the selection function over their extent. As a simple
example, consider a cluster sitting at the near edge of one cell, and
another, identical cluster sitting at the far edge of an identical cell. In
each case, the true $\delta_e$ is the same, as is
${N}_{e,{\mathrm{exp}},i}$,
but $N_e$ will be systematically smaller in the second cell, where the
cluster is more distant. Given the peculiar geometry of the cells one
is forced to construct in the LCRS, variation of the selection
function in the angular direction also can contribute to this
effect. Thus, power on scales smaller than that of the cell can
contribute to the variance in the counts-in-cells. Correcting for
these effects properly evidently requires assumptions about the
clustering on scales smaller than a cell. In Section \ref{c5_mock}, I
will show that this problem is not significant for this analysis by
considering mock catalogs of the LCRS.

\subsection{Models for the Probability Distribution Function}

In this subsection I describe models for the density distribution
function $f(\delta_e)$; in the next two subsections, I will describe
the models for the conditional probability $f(\delta_l|\delta_e)$,
which contains the information on the ``bias relation'' between the
two groups of galaxies.  For a review of the many different ways to
model the density distribution function $f(\delta_e)$, see
\citet{strauss95a}. I have tried just three: a Gaussian model,  a
first-order Edgeworth expansition, and a log-normal model. My main
results do not depend on which choice I pick, so here I will only
present results for the log-normal model, which provides the best fit
and which is the most mathematically convenient. It can be written:
%Here, I will use just three different models. The
%simplest model is a Gaussian:
%\begin{equation}
%\label{c5_gaussian}
%f(\delta_e) = \frac{1}{\sqrt{2\pi} \sigma_e}
%\exp\left(-\delta_e^2/2\sigma_e^2\right).
%\end{equation}
%This model is motivated on large-scales by the standard assumption of
%random-phase initial conditions in an inflationary universe. In this
%paper the scales I treat are not linear enough to justify the
%Gaussian model. An alternative is the Edgeworth expansion, which to
%first order is 
%\begin{equation}
%\label{c5_edgeworth}
%f(\delta_e) = \frac{1}{\sqrt{2\pi}\sigma_e }
%\exp(-\delta_e^2/2\sigma_e^2) \left[1 + \frac{1}{6} S_3
%H_3(\delta_e/\sigma_e) \sigma_e \right].
%\end{equation}
%where $H_3(\delta_e/\sigma_e)$ is the third Hermite polynomial. Note
%that this function has two free parameters: $\sigma_e$ and
%$S_3$. \citet{bernardeau94a} and \citet{juszkiewicz95a} found the
%Edgeworth expansion to be applicable to results of perturbation theory
%and $N$-body simulations, respectively, while \citet{kim98a} discuss
%the virtues of the Edgeworth expansion as a fit to the density
%distribution function of \iras\ galaxies. Finally, a good single
%parameter model for the nonlinear regime is the log-normal model:
\begin{equation}
\label{c5_lnden}
f(\delta_e) d\delta_e= \frac{d\delta_e}{\sqrt{2\pi}\sigma_e(1+\delta_e)}
\exp\left[-x_e^2/2\sigma_e^2\right].
\end{equation}
where $x_e=\ln(1+\delta_e)+\sigma_e^2/2$. \citet{blanton99b} gives
more details on various fits to the density distribution function
using this data.
% \citet{coles91a} discuss the
%observational and theoretical motivation for this model. 
%It is
%appropriate here because it accounts for the non-Gaussian tail of the
%density distribution and because it is mathematically convenient. 

%The density distributions must satisfy $\avg{\delta_e}=0$ and also be
%normalized over the range $-1\le\delta_e<\infty$:
%\begin{eqnarray}
%\int_{-1}^\infty d\delta_e \,\delta_e f(\delta_e) &=& 0;\cr
%\int_{-1}^\infty d\delta_e \,f(\delta_e) &=& 1.
%\end{eqnarray}
%The log-normal distribution automatically satisfies both these
%criteria. The peaks of the Edgeworth and Gaussian distributions must
%be shifted slightly to fulfill the first requirement; also their
%amplitude must be normalized to satisfy the second. 

\subsection{Deterministic Bias Models}

Here I describe models for the relationship between early- and
late-type density fields with no scatter. ``Deterministic'' is not
meant to refer to underlying physical principles but is simply meant
to express the fact that knowing the density of ellipticals tells you
with certainty the density of spirals, modulo Poisson noise. In this
case, the joint density distribution can be expressed as 
\begin{equation}
\label{c5_detbias}
f(\delta_l| \delta_e) = \delta^D(\delta_l - b(\delta_e))
\end{equation}
where $f(\delta_e)$ is the density distribution function of early-type
galaxies and $\delta^D(x)$ is the Dirac delta function. Under this
assumption, one can describe the models by the biasing function
$b(\delta_e)$.

The simplest model for $b(\delta_e)$ is linear bias:
\begin{equation}
\label{c5_linear}
b(\delta_e) = b_0+b_1 \delta_e.
\end{equation}
The obvious problem with this model is that if $b_1>0$, $b(\delta_e)$
can become less than $-1$. In this case, I simply set
$b(\delta_e)=0$. Another way of handling the problem is to use 
power-law bias:
\begin{equation}
\label{powerlaw}
b(\delta_e) = b_0 (1+\delta_e)^{b_1} - 1,
\end{equation}
which always remains greater than $-1$. It turns out, as I will show
below, that the power-law bias is in fact a poorer fit to the
data than linear bias in this analysis of the LCRS.

At the cost of an extra parameter, the linear bias case can be
extended trivially to quadratic bias:
\begin{equation}
\label{c5_quad}
b(\delta_e) = b_0+b_1 \delta_e + {b_2}\delta_e^2.
\end{equation}
Another possibility I try is ``broken'' bias, which is piecewise
linear with one slope in overdense regions and another in underdense
regions:
\begin{equation}
\label{c5_broken}
b(\delta_e) = \left\{\begin{array}{ccc}
b_0+b_{\mathrm{1}} \delta_e & {\mathrm{for}} & \delta_e<0 \cr
b_0+b_{\mathrm{2}} \delta_e & {\mathrm{for}} & \delta_e>0 
\end{array}\right.
\end{equation}
For each model, I require that $\avg{\delta_l}=0$, because it is meant
to represent the overdensity of late-type galaxies. In practice, this
requirement sets $b_0$, which is therefore not treated as a free
parameter in any of the above expressions.

\subsection{Stochastic Bias Models}

If variables other than the local density field are important in
determining where galaxies form, it may be that the different
formation processes of early-type and late-type galaxies cause scatter
in the relationship between their density fields.  Thus, I also
examine other models which do incorporate scatter, rewriting Equation
(\ref{c5_detbias}) by replacing the Dirac delta-function with a
Gaussian of finite width:
\begin{equation} 
\label{c5_stobias}
f(\delta_l|\delta_e) = \frac{1}{\sqrt{2\pi}\sigma_b} \exp\left[
\frac{(\delta_l- b(\delta_e))^2}{2\sigma_b^2}\right] 
\end{equation} 
where $b(\delta_e)$ and $f(\delta_e)$ are chosen as above. $\sigma_b$
parameterizes the degree of scatter in the relationship. Note that
because of the lower limit $\delta_l\ge-1$, $\avg{\delta_l|\delta_e}
\ne b(\delta_e)$ in general, although the differences will not be
important for my purposes. I do shift the peak of the Gaussian
slightly to guarantee that $\avg{\delta_l}=0$. In the case that
$f(\delta_e)$ is Gaussian, and $\sigma$ and $\sigma_b$ are small,
$f(\delta_e, \delta_l)$ reduces to a bivariate Gaussian distribution,
and the standard correlation coefficient is related to $\sigma_b$ by
\begin{equation}
r = \sqrt{1-(\sigma_b/\sigma_l)^2}
\end{equation}
Below, I will use Equation (\ref{c5_stobias}) to fit linear bias with
Gaussian scatter to the relationship between galaxy types.

% The simplest model in this case is the bivariate Gaussian
% distribution:
% \begin{equation}
% \label{c5_bivargauss}
% f(\delta_e, \delta_l) = \frac{1}{2\pi\sigma_e\sigma_l\sqrt{1-r^2}}
% \exp\left[-\frac{\delta_e^2\sigma_l^2+\delta_l^2\sigma_e^2-
% 2\delta_e\delta_l\sigma_e\sigma_l r}
% {2\sigma_e^2\sigma_l^2(1-r^2)}\right].
% \end{equation}
% The correlation coefficient
% $r=\avg{\delta_e\delta_l}/\sigma_e\sigma_l$ parameterizes the degree
% of stochasticity in the results. The case $r=1$ reduces to linear bias
% (Equation \ref{c5_linear})
% with $b_1=\sigma_l/\sigma_e$, meaning that there is no scatter in the
% relationship between $\delta_e$ and $\delta_l$. The problem with the
% bivariate Gaussian is that outliers are accorded very low probability
% and thus those that occur dominate the fit. I can handle the tails of
% the distribution more robustly by considering the bivariate log-normal
% model:
% \begin{eqnarray}
% f(\delta_e, \delta_l) &=& \frac{1}{2\pi\sigma_e\sigma_l\sqrt{1-r^2}}
% \frac{1}{(1+\delta_e)(1+\delta_l)}\cr
% & &\exp\left[-\frac{ x_e^2\sigma_l^2+
% x_l^2\sigma_e^2-
% 2\sigma_e\sigma_l r x_e x_l
% }
% {2\sigma_e^2\sigma_l^2(1-r^2)}\right].
% \end{eqnarray}
% where $x_e$ and $x_l$ are defined as in Equation \ref{c5_lnden}. This
% model retains much of the mathematical convenience of the bivariate
% Gaussian distribution. Note that the limit $r=1$ does not correspond
% to linear bias, but instead to power-law bias (Equation
% \ref{c5_powerlaw}).
% 

\section{Results from the LCRS}
\label{c5_results}

In this section, I determine the selection function of the LCRS, fit
for the density distribution function, and fit for the bias.  I then
investigate important systematic effects in the results which indicate
that the stochasticity measured from the full set of cells described
in Section \ref{c5_data} may overestimate the true stochasticity.

\subsection{Selection Function and Expected Counts}

I first fit for the luminosity function for the early- and late-type
galaxies, in each MOS field type (112-fiber or 50-fiber) and Galactic
hemisphere (north or south), denoting therefore each field as N112,
S112, N50, or S50. I limit the redshifts of the galaxies I consider to
the range where the counts-in-cells analysis will take place, from
5,000 km/s to 50,000 km/s; this limitation minimizes uncertainties due
to possible systematic effects associated with identifying higher
redshift galaxies ($> 50,000$ km/s). For details on these fits, see 
\citet{blanton99b}.

%The luminosity functions are shown in Figure \ref{c5_lcrsphi}. Note
%that the biggest discrepancies between the 50-fiber and 112-fiber
%fields are in the faintest galaxies, which is probably a result of the
%more stringent surface-brightness cut placed on the 50-fiber
%fields. As disussed above, I do not ``correct'' the 50-fiber fields to
%match the 112-fiber fields at the faint end, since the selection
%function must self-consistently include the fact that these faint
%galaxies are missing from the data.

%As an {\it a posteriori} check of the luminosity function, I plot in
%Figure \ref{c5_lcrsldsplit0} the luminosity distribution of galaxies
%as a function of redshift for the N112 galaxies of each type
%(\citealt{sandage79a}; \citealt{yahil91a}). The point of this diagram is
%to demonstrate that at each redshift, the luminosity distribution of
%observed galaxies does conform to that one would expect given the
%derived luminosity function, the stated apparent magnitude limits, and
%the galaxy density field. Note that the expected counts of galaxies
%(the solid histogram) matches closely the actual counts of galaxies
%(the dotted histogram).  The $\chi^2$ and the number of degrees of
%freedom are listed in each panel. In the range over which I fit,
%$\chi^2$ is always low. This remains true for S112, N50 and S50 as
%well. However, it is not clear whether this test of the luminosity
%function is adequate; as I will show below, even {\it severe} errors in
%the luminosity function are hard to detect with this statistic.

Figure \ref{c5_lcrsselfunc} shows as the solid histogram the expected
distribution of galaxies with redshift, assuming $\delta=0$, for all
four types of fields and for both early- and late-type galaxies. The
actual distribution of galaxies is shown as the dotted line. Note that
the two distributions follow each other reasonably well. The alert eye
will be suspicious of the apparent underdensity of actual late-type
galaxies relative to the expected late-type galaxies at low redshifts
in N112 and S112, where an large apparent overdensity exists for the
early-type galaxies. In addition, there are somewhat fewer than
expected early-type galaxies at high redshift. These features turn out
to be important, and I discuss below their ramifications.

%Similarly, I can calculate the expected counts in each cell as
%described in Section \ref{c5_data}, and take the ratio of the actual
%to the expected counts. I show the results of this comparison for each
%galaxy type in Figure \ref{c5_cells}. This figure shows the
%distribution of $N_e/{N}_{e,\mathrm{exp}}$ and $N_l/{
%N}_{l,\mathrm{exp}}$, which is in essence the data I am trying to fit
%with the model $P(N_e,N_l)$. In this figure, I have plotted the cells
%which are closer than 24,000 km/s as solid squares. They clearly have
%a small bias relative to the rest of the cells.  Furthermore, this
%result {\it persists} if I instead use the luminosity functions of
%\citet{bromley98a} in order to calculate the selection
%function.\footnote{Indeed, it was this result which first inspired my
%reanalysis of the selection function.} 
% Figure
% \ref{c5_bromley} shows the resulting cells for this exercise; it was
%in fact this figure which first inspired my reanalysis of the
%selection function. 
%I investigate this issue quantitatively in Section \ref{c5_reddep}.

%\subsection{Fitting the Density Distribution Function}

\subsection{Fitting the Bias Function}

The first task is to fit for $f(\delta_e)$ and $f(\delta_l)$, using
the method of fitting for the probability distribution function
described in Section \ref{c5_method}. The results of fits to a
log-normal distribution are shown in Table \ref{c5_biasrd}.
%A comparison of the results of
%Gaussian, Edgeworth, and log-normal fits is shown in Table
%\ref{c5_pdffits}; likelihoods are given relative to the best fit
%Gaussian model. The resulting distributions $P(N_e)$ and $P(N_l)$ are
%shown in Figure \ref{c5_pdf}.  For comparison, the likelihood function
%of each fit is shown in Figure \ref{c5_pdflike}; I also show in that
%figure the bootstrap errors, to show that they correspond to the width
%of the region in which ${\mathcal{L}} < {\mathcal{L}}_{\mathrm{min}} +
%1$.  From the likelihoods listed in Table \ref{c5_pdffits} and the
%comparison of the distributions in Figure \ref{c5_pdf}, it is clear
%that the Gaussian distribution is a poor approximation. The log-normal
%is slightly better than the Edgeworth distribution, and for that
%reason, as well as its simplicity, I use it throughout. 
It is already clear from the results in this table that the early-type
galaxies are more clustered than the late-type galaxies by about
20--30\%. 
%Furthermore, it is approximately true that
%$\sigma_eS_{3,e}\approx\sigma_lS_{3,l}$, consistent with approximately
%linear bias between the two fields (\citealt{fry93a}). However, using
%the maximum likelihood method presented here for fitting for the bias
%provides a considerably more powerful test of linearity, as shown
%below.

\begin{figure}
\figurenum{\fignum}
\plotone{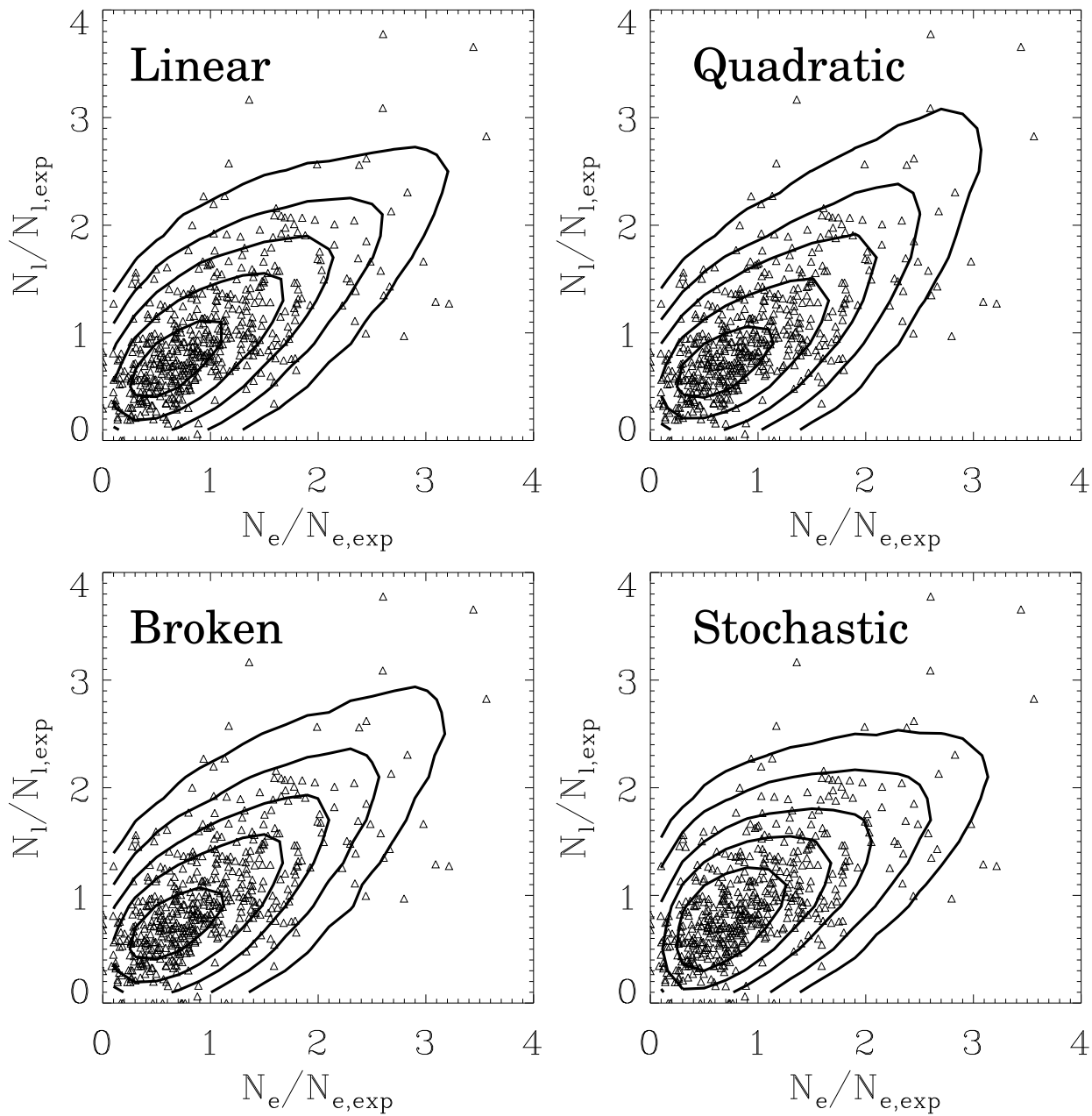}
\caption{\label{c5_contours} Actual joint counts-in-cells for early- and
late-type galaxies {\it (triangles)} and model joint distribution
{\it (contours)}, based on the fitted probability distribution
function and bias relation convolved with Poisson noise for the set of
cells in the LCRS. From the inside out, the contours include 30\%,
70\%, 85\%, 93\%, and 97\% of the model cells.
Results are shown for four different forms of the
bias relation, as labeled. I use the log-normal density distribution
for $f(\delta_e)$.}
\end{figure}

With the parameters of the $f(\delta_e)$ distribution in hand, I now
fit for the bias relation $f(\delta_l|\delta_e)$.  I try all of the
models described in Section \ref{c5_method}, as listed in the top
section of Table \ref{c5_biasrd}.  For the linear fit I find
$b_1=0.76\pm 0.02$; that is, the late-type galaxies are underbiased
with respect to the early-type galaxies, to a degree which agrees with
the conventional wisdom.  The power-law fit is worse than the linear
fit, and I will not consider it further. Some nonlinearity is detected
in the quadratic bias case, at rather high significance (6$\sigma$;
although read below for words of caution in interpretation), in the
sense that the slope of the bias steepens at large densities. The
broken bias model also shows this effect, at somewhat lower
significance.  Stochasticity is detected at about 10$\sigma$
significance, with $\sigma_b=0.21\pm 0.02$.  Contours showing the
model distributions of $P(N_e,N_l)$ superposed on the data are shown
in Figure \ref{c5_contours} for the linear, quadratic, broken, and
stochastic models. Notice that the quadratic and broken models curve
upward relative to the linear model to fit the data better, while the
stochastic model has fattened contours because of the scatter it adds
to the relation.

\begin{figure}
\figurenum{\fignum}
\plotone{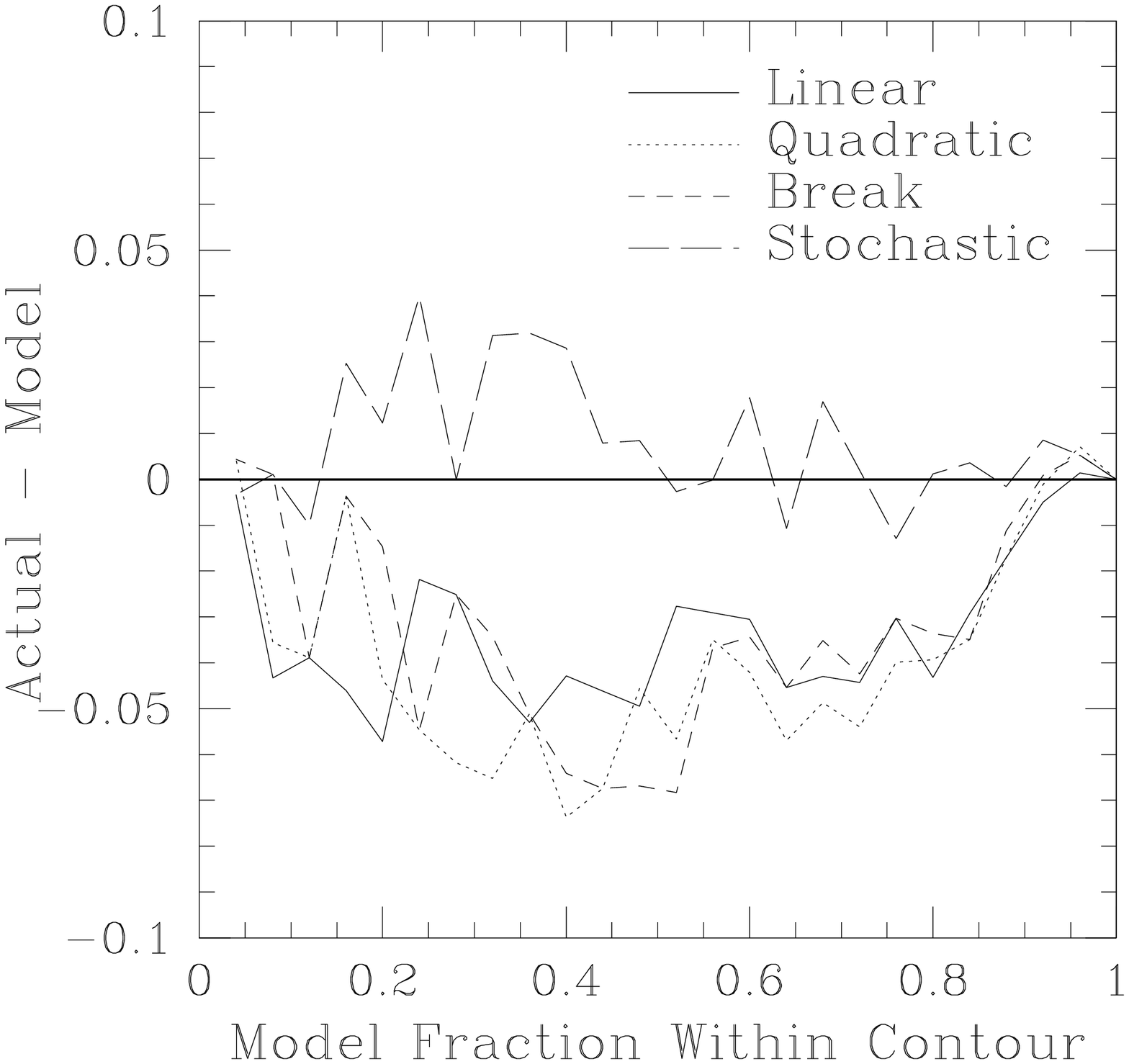}
\caption{\label{c5_ks} Difference between the model fraction of cells
within the contours in each panel of Figure \ref{c5_contours} and the
actual fraction of cells within those contours. }
\end{figure}

One can investigate the difference between these contour plots a bit
more quantitatively. I do so by finding a set of contours of the model
probability which contain an increasing fraction of the model cells,
from $0$ to $1$. I can compare this fraction to the fraction of actual
cells which are contained in each contour. The difference between the
fraction of model cells to the fraction of actual cells is plotted for
each model in Figure \ref{c5_ks}. It is clear that the model which
follows the probability contours most closely is the stochastic
model. A refinement of this analysis would be to compare each cell to
the model $f(\delta_e,\delta_l)$ convolved with the Poisson noise of
that cell alone, and to calculate a
two-dimensional statistic analogous to the Kolmogorov-Smirnov test
(\citealt{lupton93a}).

The likelihood functions for each type of fit are shown in Figure
\ref{c5_biaslike}.  This
plot shows that bootstrap gives errors comparable to those calculated
from the likelihood function.  The likelihoods relative to the linear
fit are also shown in Table \ref{c5_biasrd}, and indicate that the
stochastic linear bias is clearly the best model I have considered. I
have compared each of the two-parameter fits to the linear fit by
using a likelihood ratio test, whose results I show in the last column
as the probability $P_{\mathrm{linear}}$ of getting the observed
likelihood difference if the true bias relation were linear.  Note
that since I only ran 200 realizations for each estimate, there is a
lower limit on $P_{\mathrm{linear}}$ of 0.005. From these results, it
is clear that I detect nonlinearity at a statistically significant
level, and stochasticity at an extremely significant level. 

Are stochasticity and nonlinearity both present?  To address this
question, rather than fitting a nonlinear and stochastic model, which
would be complicated and would introduce an extra parameter, I instead
ask: is the detected nonlinearity consistent with what one would find
if the stochastic, linear model was correct? For example, in a
realization of the stochastic, linear model, it can happen that the
few high-density cells happen to all scatter below the mean
$\delta_l=b \delta_e$. In this case, a nonlinear fit will have a
higher likelihood than a linear fit; it will not, on the other hand,
be correct. Can the degree of improvement of the nonlinear fits over
the linear fits be simply attributed to this effect?  I test this by
looking at the likelihood ratio between the quadratic fit and the
linear fit in a set of Monte Carlo realizations of the {\it
stochastic} bias model. This experiment shows that the likelihood
ratio is on average ${\mathcal L}_{\mathrm{linear}}-{\mathcal
L}_{\mathrm{quad}}\approx-11\pm 9$. That is, if there is in reality
some scatter around linear bias, the data always is better fit by the
deterministic quadratic model than by the deterministic linear model,
and the measured quadratic parameters are meaningless. In fact, since
the measured likelihood ratios between the nonlinear and linear fits
are in this range, I cannot truly claim to have measured nonlinearity
here.

While these likelihood ratio tests reveal the relative quality of the
fits, one can evaluate the absolute quality by looking at the
likelihood distribution of fits to a set of Monte Carlo realizations
of each model, and calculating the probability $P_{\mathrm{random}}$
of getting a lower best-fit ${\mathcal{L}}$ for the model; if this
probability is near unity, then the data do not fit the model as well
as they should according to the realizations, whereas if it is less
than $\sim 0.5$ the model fits the data as well as could be
expected. This method is not ideal, but it gives a rough calibration
of whether the fit is decent. From the top section of Table
\ref{c5_biasrd}, it appears that the only model that does a reasonable
job of fitting the full set of cells is the stochastic model.

The level of stochasticity detected here corresponds to $r\approx
0.87$; for comparison, the moments method of TB99 would estimate $r$
for this counts-in-cells distribution to be $r = 0.73 \pm 0.01$.
Apparently a considerable amount of the ``stochasticity'' measured by
the moments method is due to an inadequate characterization of the
distribution of the densities, most likely because the method does not
account for the non-Gaussianity of the Poisson distribution at low $N$
or for the lower limit on the overdensity fields of $-1$. On the other
hand, as discovered in the next section, even this small level of
measured stochasticity may be fictitious, due to systematic errors in
the selection function.

\subsection{Testing for Systematic Errors}
\label{c5_system}

In order to probe the susceptibility of these results to systematic
errors, I have run a battery of tests. First, I
briefly list a number of tests which made no significant difference in
my main results. Second, I describe the effects of the central
magnitude cut.
%; third, I describe the effects of changing the
%configuration of the cells; fourth, I show how errors in the
%luminosity function can affect the results. 
Finally, in the next subsection, I show that there is a
redshift-dependence of the results.

A number of tests have very little effect on the results concerning
stochasticity. For instamce, whether or not $\delta_e$ or $\delta_l$
is used as the independent variable has no effect on the results. In
addition, identical results are found in the Northern and Southern
Galactic hemispheres. Furthermore, there is no dependence of the
stochastic fit on changes in: the form of the density distribution
function used, the $K$-correction applied, the completeness correction
applied, or the cell configuration used. (The nonlinear fits {\it do}
depend somewhat on cell configuration, but I argued above that the
nonlinear fits in this case were probably meaningless). A more
complete discussion of these issues can be found in
\citet{blanton99b}.
%First, I have rerun the analysis, using the late-type
%galaxy density as the independent variable and the early-type galaxy
%density as the dependent variable. These results
%%, listed in Table \ref{c5_reverse},
%are consistent with simply being the inverse of the results in Table
%\ref{c5_biasrd}, so there is no systematic effect associated with
%which variable is dependent and which is independent.  
%%To compare the
%%stochasticity, I compare $\sigma_b/\sigma_l$ in Table \ref{c5_biasrd}
%%to $\sigma_b/\sigma_e$ in Table \ref{c5_reverse} (both are $\approx
%0.5$), since $\sigma_b$ scales approximately as the standard deviation
%of the dependent variable.  \
%Second, I have split the cells by Galactic
%hemisphere; 
%as I show in Table \ref{c5_NS}, 
%the results for each
%hemisphere are consistent with each other and with the results in
%Table \ref{c5_biasrd}.  Third, using the Gaussian or Edgeworth density
%distribution functions instead of the log-normal model had no
%substantial effect on the results.  Fourth, I have tried completely
%neglecting the incompleteness corrections $f_g(m)$ (though retaining
%$f_{t}$); this omission has no effect on the results.  Fifth, I tried
%both neglecting $K$-correction entirely and using different
%$K$-corrections for the early- and late-type galaxies
%(\citealt{frei94a}; I used their elliptical results for the
%early-types and their Scd results for the late-types). Neither
%alteration to the form of the $K$-correction had any effect on the
%results.

A suspicious element of the selection of galaxies in the LCRS which I
address here is the central magnitude cut.  Because
aperture magnitudes of fixed angular size exclude a varying fraction
of galaxy light depending on redshift,
%and
%because galaxies do not have flat profiles, 
the central magnitude defined by the LCRS is actually a redshift
dependent quantity. The sense is that at low redshift, less of the
total galaxy flux is contained within a central angular radius, and
thus $m_c$ is likely to be higher. Thus, this cut will preferentially
exclude low-redshift galaxies. An aperture magnitude with a diameter
of 3--4$''$, as used in the LCRS, depends much more strongly on
redshift for an exponential profile (characteristic of late-type
galaxies) than a de Vaucouleurs profile (characteristic of early-type
galaxies) in the redshift range under consideration here.  It is
therefore conceivable that late-type galaxies are being preferentially
excluded at low redshifts, causing an apparent low density relative to
the densities of early-type galaxies in that regime.
%, as demonstrated in Figures
%\ref{c5_lcrsselfunc} and \ref{c5_cells}.
%, and \ref{c5_bromley}.

However, the situation appears not to be that simple, as I show by
performing the following experiment. Instead of using the central
magnitude limit $m_{c,\mathrm{cut}}$ (as defined in Equation
\ref{c5_mcdef}) used by the survey, which excludes 12\% of the
galaxies, I enforce a somewhat more stringent limit in $m_c$ (by about
0.2 magnitudes) which excludes about 24\% of the galaxies. If the
effect descibed in the previous paragraph is important, one would
expect the estimated galaxy density field to change
significantly. However, it does not. 
%To understand why, consider
%Figure \ref{c5_lfmccut}, which shows the luminosity function in the
%N112 fields for the full sample as the solid line and for my more
%stringent sample as the dotted line, for each galaxy type. Clearly the
%cut eliminates a number of intrinsically faint, late-type galaxies;
%faint galaxies in the sample are preferentially excluded because they
%are nearby. 
To understand why, consider Figure \ref{c5_deltaz}. The top panel in each
column shows the ratio of the number of galaxies per unit redshift in
the stringent sample to the number in the full sample, for the N112
fields, $N_{\mathrm{stringent}}(z)/N_{\mathrm{full}}(z)$, for each
galaxy type. Clearly, as described in the previous paragraph, the
late-type galaxies at low redshift are preferentially
excluded. However, as shown in the middle panel, the selection
function for the stringent sample (shown as the dotted line) also
changes relative to the full sample (shown as the solid line). This
happens because the average number of faint galaxies relative to
bright galaxies is underestimated for the stringent sample. Since
faint galaxies are not observable at large distances, this change does
not affect the selection function at large redshifts. Indeed, as the
bottom panel shows, the ratio $N(z)/N_{\mathrm{exp}}(z)$ for the
stringent sample is almost identical to that of the full
sample. Correspondingly, 
%as shown in Table \ref{c5_mccuttable}, 
the results of the density distribution and bias fits are unchanged as
well. Thus, the effect of the central magnitude cut seems not to be
crucial.

\subsection{Redshift Dependent Selection Effects}
\label{c5_reddep}

\begin{figure}
\figurenum{\fignum}
\plotone{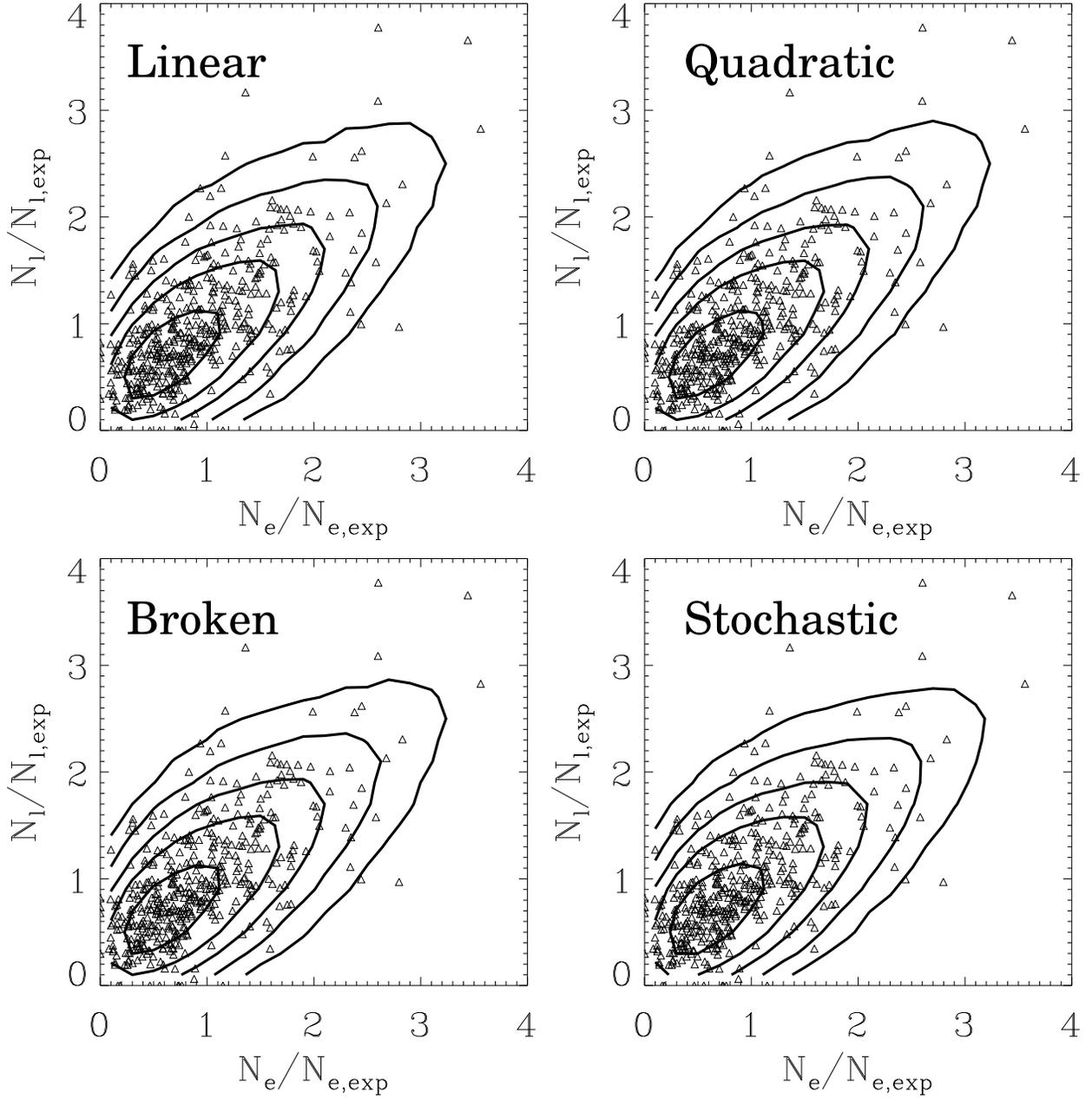}
\caption{\label{c5_contours_nir2} Same as Figure \ref{c5_contours},
excluding low-redshift cells ($cz < 24,000$ km/s). Notice how similar
all the model distributions are.}
\end{figure}

\begin{figure}
\figurenum{\fignum}
\plotone{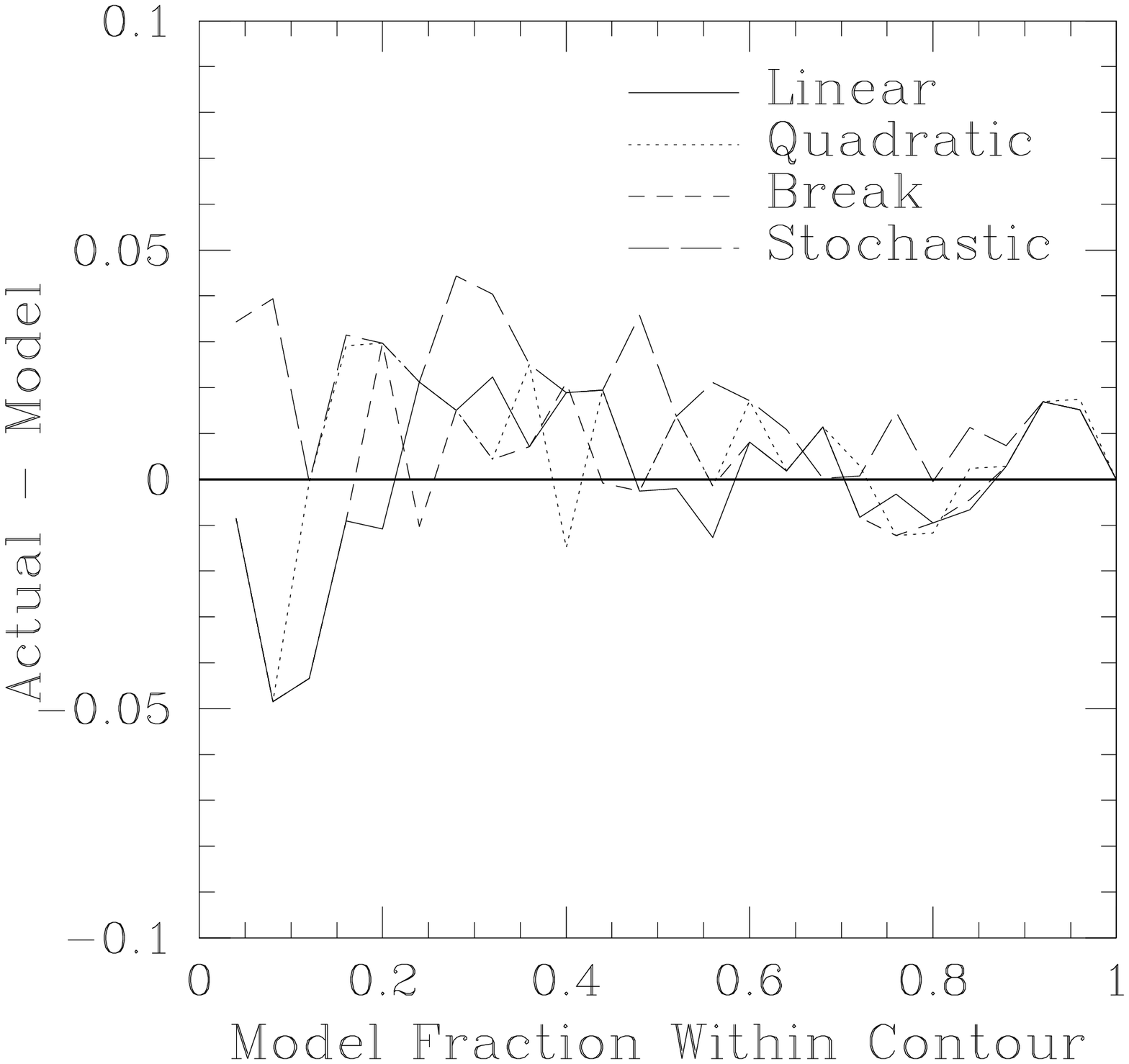}
\caption{\label{c5_ks_nir2} Same as Figure \ref{c5_ks}, excluding
low-redshift cells ($cz < 24,000$ km/s).}
\end{figure}

In order to demonstrate the redshift-dependence of the results, I 
%split the
%cells into a high-redshift half and a low redshift half (still using,
%however, the selection function determined for the whole sample). As
%shown in the third and fourth sections of Table \ref{c5_biasrd}, I
%find that the low redshift half retains the large amount of
%nonlinearity and stochasticity, while the high redshift half has no
%detectable signal.
%I can be more precise and 
cut the two innermost rings of cells out of the sample,
and fit to the rest, as shown in the bottom section of Table
\ref{c5_biasrd}. This set of cells shows no nonlinearity, and a much
reduced stochasticity. The contours showing the model $P(N_e,N_l)$
superposed over the data for this set of cells are shown in Figure
\ref{c5_contours_nir2}. Again, I compare the model to actual fraction
of cells within each contour in Figure \ref{c5_ks_nir2}, finding this
time no apparent difference between any of the models. I can again
express the stochasticity in terms of the correlation coefficient
$r\approx 0.95$; the moments method of TB99 obtains $r = 0.93 \pm
0.03$ for this set of cells.  This result indicates that most of the
signal for stochastic bias was coming from the two innermost rings of
cells. 

This dominance of the inner two rings cannot be ascribed to the higher
signal-to-noise ratio of these cells. To demonstrate this fact, I perform
Monte Carlo realizations using all the cells except the two inner
rings, but using the parameters for linear stochastic bias determined
using all of the cells ({\it i.e.}, using the parameters in the top
section of Table \ref{c5_biasrd}).  I find that the maximum likelihood
fit detects $\sigma_b\approx 0.2$ (the correct value for the tests)
with a likelihood difference with respect to the linear fit between
$-40$ and $-80$, not $-8$ as for the real data.  Nor does the redshift
dependence appear to be a result of luminosity-dependent bias (in the
sense that fainter galaxies, observable only at low redshift, have a
different relative bias between galaxy types). I show this by
performing the analysis again, using only galaxies brighter than
$M=19.3$, the faintest galaxy luminosity observable at 24,000 km/s; I found
little change in the results.
%, as shown in Table \ref{c5_bright}.

\begin{figure}
\figurenum{\fignum}
\plotone{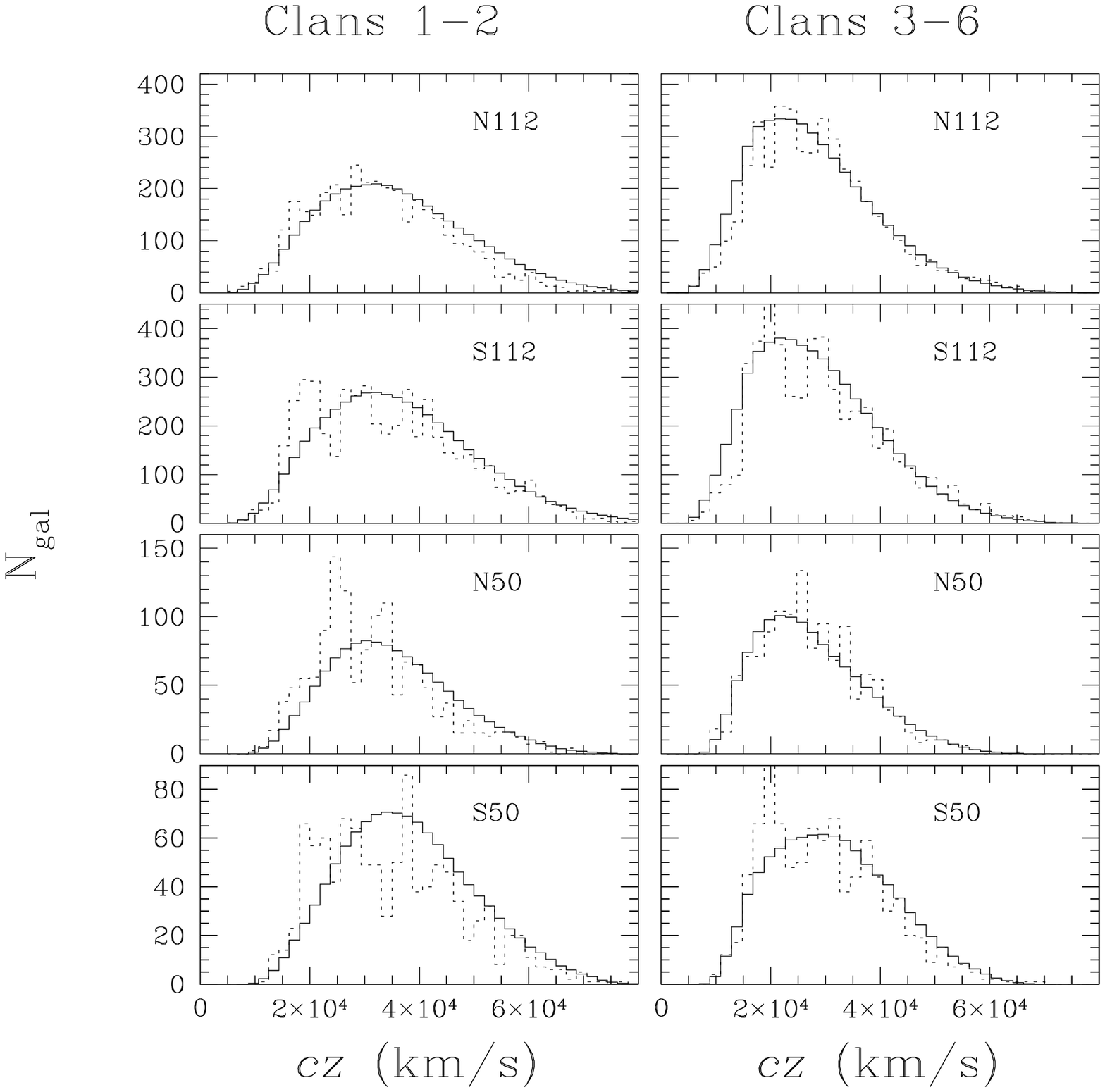}
\caption{\label{c5_lcrsselfunc} Actual galaxy redshift distribution
{\it (dotted line)} and expected redshift distribution {\it (solid
line)}, based on the luminosity function and the flux limits. Results
are shown for early and late type galaxies in each sample, as
labeled. Note that at low redshift, the early-type and late-type
density fields are noticeably different.}
\end{figure}

\begin{figure}
\figurenum{\fignum}
\plotone{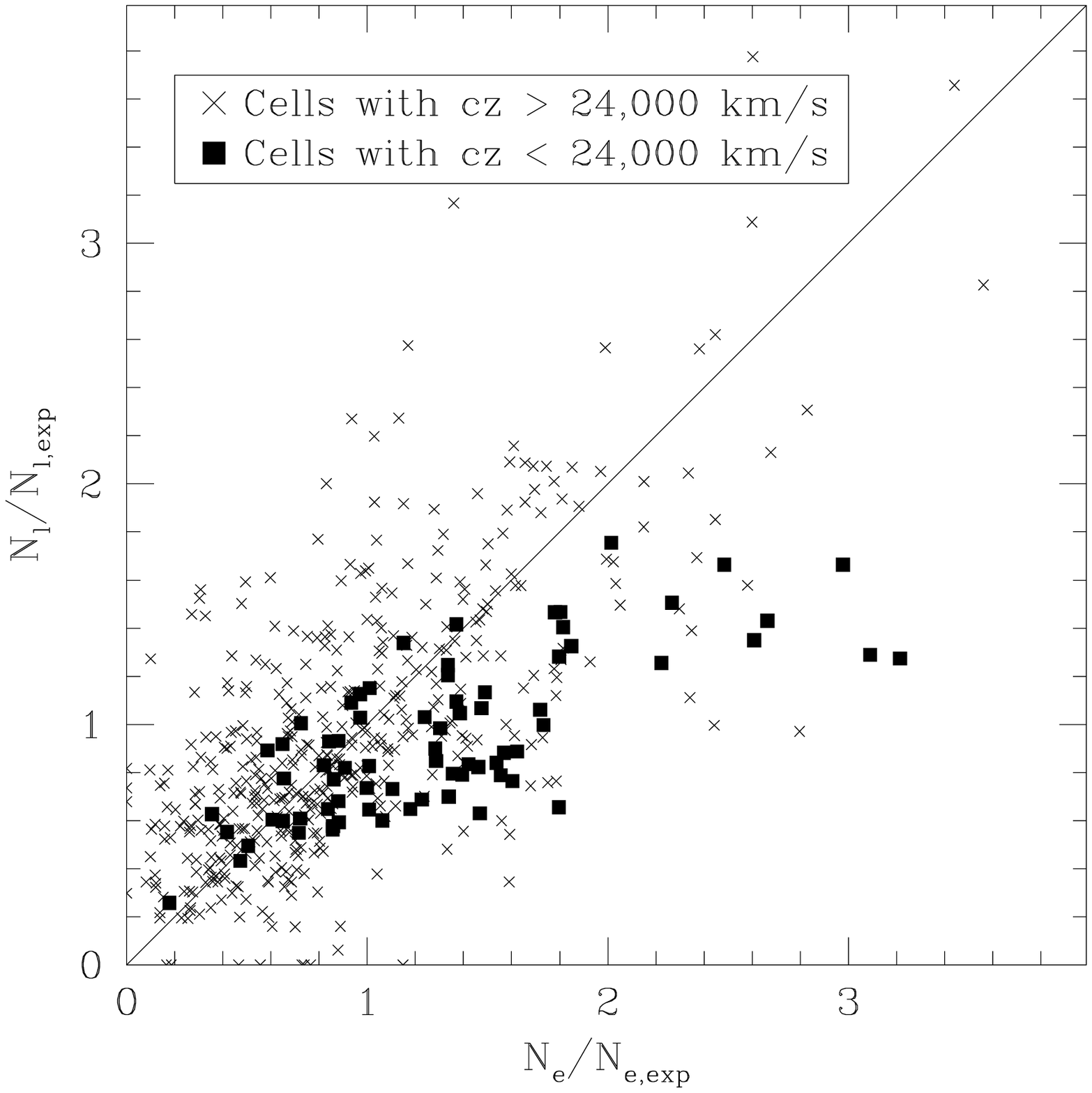}
\caption{\label{c5_cells} Joint overdensity distribution of early-type
galaxy {\it (x-axis)} and late-type galaxy {\it (y-axis)}
counts-in-cells for the LCRS. The low-redshift cells are shown as
solid squares and the high-redshift cells are shown as crosses. Note
that it appears as if the low-redshift cells are systematically less
biased than the high redshift cells, which accounts for the increased
stochasticity and nonlinearity when these cells are included. This low
bias could occur if at low redshifts the selection function was
underestimated for the early-type galaxies or overestimated for the
late-types.}
\end{figure}

In order to understand the effect better, consider Figure
\ref{c5_cells}. This figure shows the distribution of
$N_e/{N}_{e,\mathrm{exp}}$
and $N_l/N_{l,\mathrm{exp}}$ among the cells.
I have marked the cells in the inner two rings
with square boxes and the other cells as crosses.  It is clear that
the nearby cells have a different distribution than the rest of the
cells. What this indicates is that the selection function is either
overestimated for late-type galaxies or underestimated for early-type
galaxies at low redshifts, which is plausible on consideration of
Figure \ref{c5_lcrsselfunc}. I have performed the same analysis using
selection functions based on the luminosity functions of
\citet{bromley98a}
%, as shown in Figure \ref{c5_bromley}, 
and find the
same effect.

\begin{figure}
\figurenum{\fignum}
\plotone{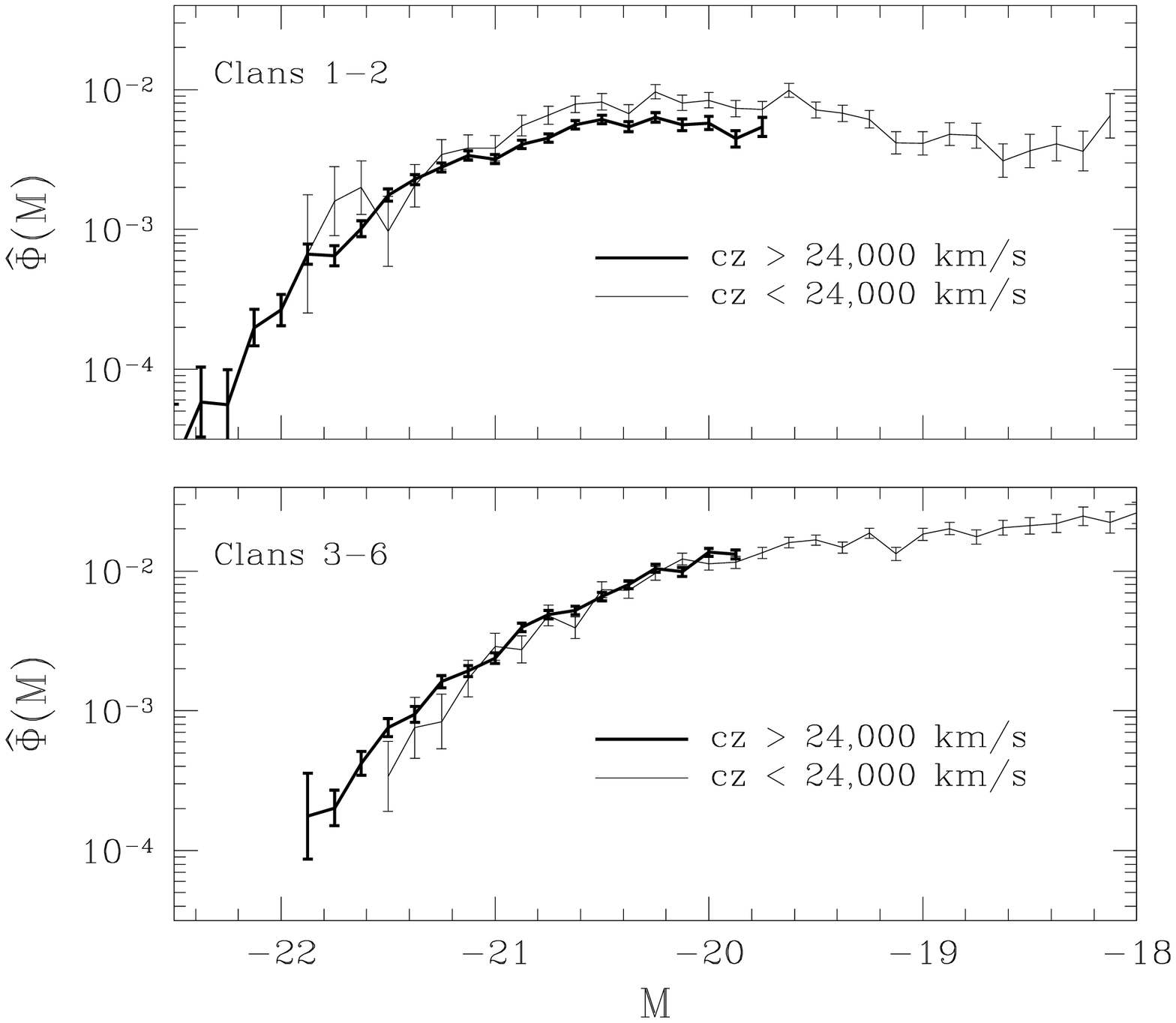}
\caption{\label{c5_lfhilo} The luminosity function $\hat\Phi(M)$, as 
defined in Equation \ref{c5_hatphi}, for LCRS galaxies of Clans 1 and 2
{\it (top panel)} and of Clans 3 through 6 {\it (bottom panel)} in the
N112 sample. I show fits separately for high redshift and low redshift
subsamples. Note that for late-type galaxies the two samples are
consistent, while for early-type galaxies there is a normalization
error of about 40\%, in the sense that early-type galaxies are missing
at high redshift.}
\end{figure}

\begin{figure}
\figurenum{\fignum}
\plotone{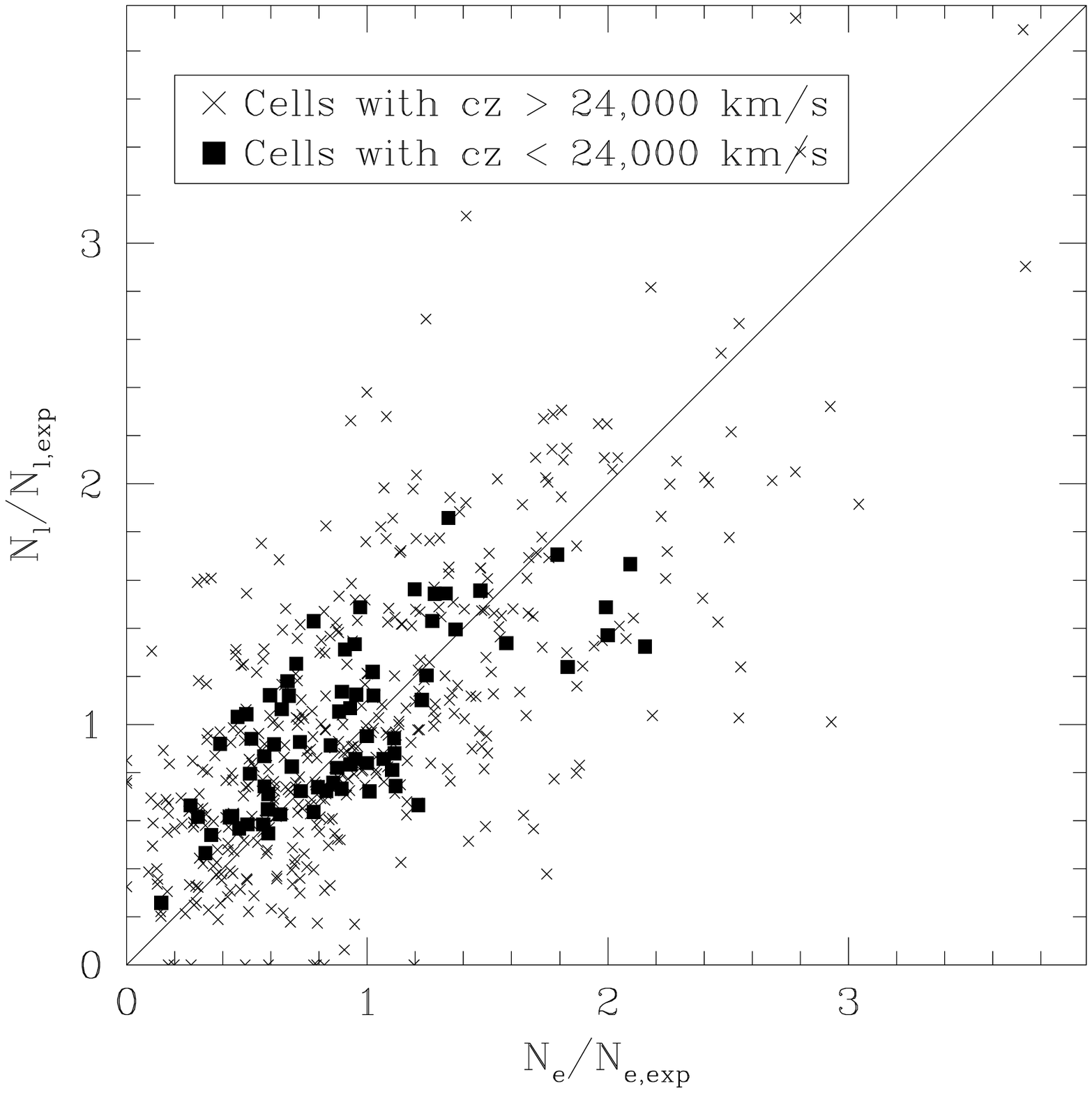}
\caption{\label{cells.lohi2} Same as Figure \ref{c5_cells}, but
using the low-redshift luminosity functions from Figure
\ref{c5_lfhilo} for the inner two rings of cells, and the
high-redshift luminosity functions for the outer cells. Notice how the
distribution of counts in low-redshift cells is now much more
consistent with the rest of the distribution.}
\end{figure}

In Figure \ref{c5_lfhilo}, I show the luminosity function of the
galaxies in the N112 sample again, this time fitting separately to the
high-redshift portion ($cz>24,000$ km/s) and the low-redshift portion
($cz<24,000$ km/s).  Clearly, there are significantly fewer early-type
galaxies detected at high redshift than at low redshift. Given the
relatively shallow depth of the survey, it is not likely that that
this difference is due to evolution of the population of early-type
galaxies.  Thus, the large apparent overdensity of early-types in the
low-redshift region (see Figure \ref{c5_lcrsselfunc}) might exist only
because the normalization is underestimated due to early-type galaxies
being missed at high redshifts. In this case, the observed
stochasticity could be due simply to fluctuations in the apparent
overdensity field caused by errors in the selection function. I test
this by fitting for the bias in all cells, but determining the
expected counts using the low-redshift luminosity function for the
inner two rings of cells, and the high-redshift luminosity function in
the rest of the cells. The results are listed in the second section of
Table \ref{c5_biasrd}. The linear fit is basically unchanged. The
nonlinear fits change dramatically, though as I showed above, such
changes should not be too surprising. Finally, the stochasticity is
reduced significantly, to $0.16\pm 0.02$. Remembering that $\sigma_b$
adds quadratically, this result indicates that a large proportion of
the scatter was indeed simply due to the redshift dependence. Figure
\ref{cells.lohi2} shows the joint counts-in-cells that these fits were
based on, showing (as in Figure \ref{c5_cells}) the low redshift cells
separately from the high redshift cells. One can see clearly that the
estimated early-type densities of the low redshift cells are quite
different from those in Figure \ref{c5_cells}, and that the
distribution of low redshift cells now appears consistent with the
distribution of the rests of the cells.

The question remains where this redshift dependence comes from.  I
have already shown that the central magnitude selection criterion does
not affect the results; in any case, one would expect it to
preferentially exclude late-type galaxies at low redshift, since
late-types are more extended than early-types. Similarly, the use of
isophotal magnitudes, which depend on redshift due to $(1+z)^4$
surface-brightness dimming, would also preferentially exclude
late-type galaxies. In addition, I show below using mock catalogs that
the use of isophotal magnitudes does not affect the results. Another
possibility is that the spectral classification scheme is
misclassifying galaxies.  However, in the case of misclassification,
one would expect to see comparable errors in the normalization of both
galaxy types, not just the early-types. A final, speculative
possibility is that early-type galaxies at high redshift tended not to
be be identified as galaxies in the survey in the first place, since
they are compact enough to look like stars. I am currently looking at
the redshift distribution of galaxies and issues of galaxy selection
using the superior imaging and spectroscopic data of the SDSS
(\citealt{gunn95a}), and it is possible that this work will lend
understanding to the problems faced in the LCRS.

Until the nature of the galaxy selection in this survey is more fully
understood, I recommend taking most seriously the results in the
bottom section of Table \ref{c5_biasrd}, which exclude the two
innermost rings, and indicate a bias which is linear, with perhaps
some mild scatter, and an amplitude of $b_1\approx 0.8$.

\section{Results from Mock Catalogs}
\label{c5_mock}

Because of the peculiar geometry and selection effects of the LCRS, it
is necessary to test this method against mock catalogs where I have
simulated all of the properties of the survey. I would also like to
test whether some of the systematic trends with redshift found in the
last section can be explained by selection effects in the survey.

\subsection{Simulations}

For current purposes I run particle-mesh simulations of a 300 $h^{-1}$
Mpc box using $256^3$ particles and $512^3$ grid cells, using a code
provided by Renyue Cen.  I use the flat CDM model with $\Omega_m=0.4$
and $\Omega_\Lambda=0.6$; the angular diameter distances and the
distance moduli for the mock catalogs are calculated using this model,
although the analysis of the mock surveys are performed using the
$\Omega_0=1$ model for the redshift-distance relation, as they are for
the real survey.  To select the late-type galaxies, I simply pick dark
matter particles at random. To select the early-type galaxies, I
smooth the density field with a 3 $h^{-1}$ Mpc Gaussian filter, and
apply a threshhold of $\delta_{c}=0.25$, at $z=0$; every dark matter
particle above the threshold has an equal probability of becoming an
early-type galaxy.  Mock catalogs are drawn from three realizations of
this model.  Note that since the box size is smaller than the redshift
limits of the survey, I must extend the box periodically in each
direction in order to simulate the LCRS. 

As a benchmark against which to compare the mock catalogs, I take the
simulation and divide it into cubic cells 25 $h^{-1}$ Mpc on a side. I
subsample the galaxies such that there are about 20--30 galaxies of
each type in each cell. I refer to this sample as the {\it
benchmark catalog}. It is free of all of the selection effects
associated with the real survey, as well as redshift-space
distortions.  The cells are equivalent in volume to the cells
described in Section \ref{c5_data}. I have listed in the top section of
Table \ref{c5_benchmark} the results of fitting the bias for all such
cells for all three realizations simultaneously. I will evaluate the
degree to which the selection effects affect my results by comparing
realistic mock catalogs to the results for these benchmark cells.

To create realistic mock catalogs, I pick a random particle in the
simulation to represent the observer.  In order to evaluate the
effects of the cell shapes independent of other selection effects, I
create a catalog using the angular and redshift limits of all of the
cells, without regard to flux limits. I refer to this catalog as a
{\it volume-limited catalog}. For this catalog, I do implement
redshift-space distortions, although these do not make a significant
difference in the results. Again, I subsample the galaxies such that
there are approximately 20--30 galaxies of each type in each cell.

To create flux-limited catalogs, I assign each galaxy an absolute
magnitude randomly from the luminosity function determined in Section
\ref{c5_results} for N112 galaxies, depending on which type of galaxy
it is. I then ``observe'' the galaxies in the simulation box, using
the angular and photometric limits of the LCRS
(\citealt{shectman96a}). For most of the catalogs, I assume the
observers have the ability to measure total magnitudes, and apply only
the apparent magnitude limits, not the central magnitude limits. I
explore below the effects of using isophotal magnitudes and
implementing the central magnitude limits. I create two types of
flux-limited mock catalogs: first, {\it fully-sampled catalogs}, in which
I take the redshift of every galaxy within the flux-limits of each
field; second, {\it undersampled catalogs}, in which I select the targets
in each field based on the number of fibers available for that field
(allowing for about 5\% of the fibers to be accidentally placed on
stars). Furthermore, for the undersampled catalogs, there is a
probability of failing to observe the galaxy which is a function of
its magnitude, given by $f_g(m)$. In the real LCRS, fibers could not
be placed more closely than $55''$; I implement that restriction in
the undersampled catalogs as well.  Finally, in accordance with the
stated photometric errors in \citet{shectman96a}, I included 1$\sigma$
magnitude errors of $0.1$ for $m<17$ and $0.17$ for $m>17$, as well as
1$\sigma$ redshift errors of $67$ km/s.  
%An example of one of the
%undersampled catalogs is shown in Figures \ref{c5_catmocke} and
%\ref{c5_catmockl}.

In order to determine the effects of cosmic scatter and to test
whether the survey constitutes a fair sample, I draw thirty
undersampled mock catalogs from three realizations. After fitting for
the bias in each catalog, I compared the standard deviation of the
results to the estimated errors. They were almost identical,
indicating that the errors due to cosmic scatter are no bigger than
the other statistical errors.

\subsection{Analysis of the Mock Catalogs}

The results of these mock catalogs are listed in Table
\ref{c5_benchmark}.  First, I compare the benchmark catalog, which
consists of all three realizations divided into cubical cells, to the
volume-limited catalog, which uses cells of the same shape used in the
actual survey. The benchmark catalog has much smaller error bars
because it probes considerably more volume than the other
catalogs. The fluctuation amplitudes $\sigma_l$ and $\sigma_e$
measured for the volume-limited catalog are significantly smaller than
for the benchmark catalog, indicating that the cell shapes in the LCRS
probe effectively larger scales than cubes of equivalent
volume. However, the bias fits change very little between the two
catalogs. No parameter changes more than about 1.5$\sigma$. Note that
the bias implemented here is only slightly scale-dependent, and that
if it were strongly scale-dependent, the difference between the
volume-limited and benchmark catalogs might be larger, because the two
catalogs probe somewhat different scales.

Second, I consider the fully sampled catalog, which implements the
flux-limits of the survey, but not the finite number of fibers or the
fiber collision effects. The differences in all the parameters is
quite small. $\sigma_e$ and $\sigma_l$ do increase {\it slightly} by
about 1.5$\sigma$; in addition, the stochasticity in the bias
$\sigma_b$ also increases by almost 1.5$\sigma$, but it remains much
smaller than that measured in the data.  It is possible that these
increases are due to the variation of the selection function across
the cells, as explained in Section
\ref{c5_method}. However, this effect is apparently too small compared
to the noise to be important for the LCRS, though it may be of concern
to larger surveys if they strive for more precision.

Third, the results of the undersampled catalog, which implements the
effects of a finite number of fibers and fiber collisions, are again
almost identical to the fully-sampled case. $\sigma_e$ and $\sigma_l$
are reduced by about $1\sigma$ apiece, which could just as easily be
due to chance as it could be due to the sampling effects. The only
large change is in $\sigma_b$, which does decrease rather
substantially from the fully-sampled case.

%Fourth, I have implemented different cell shapes for the mock catalogs
%as well, using the ``Contiguous'' and ``Shallow'' configurations
%described in Section \ref{c5_data}. As in the case of the data, the
%nonlinear bias fits suffer most greatly under these changes, though
%the changes in the mock catalogs are not nearly as great as for the
%real data. Perhaps the real universe has somewhat more stochasticity
%in the bias or perhaps more scale-dependent bias than the mock
%catalogs do, which could be causing greater changes when cell shapes
%change.

Given the results in this section, I conclude that the selection
function, variable flux limits, finite sampling, and fiber collisions
do not affect the results significantly. 

\subsection{Isophotal and Central Magnitude Limits}

The catalogs I examine above are purely flux-limited and assume
observers can measure total magnitudes.  However, I also want to probe
the effects of using isophotal magnitudes, as well of implementing the
central magnitude cut. In order to do so, I must also model the
surface brightness profiles and characteristic radii of the
galaxies. I adopt a very simple picture here, since I am interested
not in a perfect model of the galaxy distribution but only in some
estimate of how adding these realistic observational effects changes
one's estimate of the galaxy density field. I model the early-type
galaxies as pure de Vaucouleurs profiles:
\begin{equation}
I(R) \propto
\exp\left\{-7.67\left[\left(R/R_{\mathrm{deV}}\right)^{1/4}-1\right]
\right\},
\end{equation}
where $R$ is the distance from the center of the galaxy and
$R_{\mathrm{deV}}$ is a characteristic scale length. I model the
late-type galaxies as bulge components with de Vaucouleurs profiles,
which characteristic scale $R_{\mathrm{bulge}}$, plus disk components
with exponential profiles:
\begin{equation}
I(R) \propto \exp\left[R/R_{\mathrm{disk}}\right],
\end{equation}
where again $R_{\mathrm{disk}}$ is a characteristic scale length. For
these galaxies I fix $R_{\mathrm{bulge}}/R_{\mathrm{disk}}=0.6$ and
$B/T=0.4$ (see \citealt{binney98a} for the definition of $B/T$), which
are appropriate choices for Sbc galaxies (\citealt{kent85a}). In order
to determine the scale lengths, I follow \citet{sodre93a} and write:
\begin{equation}
\log_{10} \left[R / R_0\right] = - A
(M + 20) + \epsilon(\sigma_R).
\end{equation}
where $\epsilon(\sigma_{R})$ represents Gaussian noise with a standard
deviation $\sigma_{R}$.  In accordance with semi-analytic models I set
$A=0.13$ (\citealt{dalcanton97a}). Furthermore, I choose
$\sigma_{R,\mathrm{deV}}=0.13$, $\sigma_{R,\mathrm{disk}}=0.3$,
$R_{{\mathrm{disk}},0}=2$ $h^{-1}$ Mpc, and $R_{{\mathrm{deV}},0}=3$
$h^{-1}$ Mpc. These parameters are not unique; they simply produce a
distribution of $m$ and $m_c$ which is reasonably like the data.  I
tried several variations on these parameters as well, with no
significant change in the results.  The profile of each galaxy is
scaled to the appropriate redshift and convolved with the seeing,
which for simplicity I assume to be Gaussian with a FWHM of $\sim
1.8''$. Using a Gaussian seeing profile is somewhat unrealistic; it
will make very little difference to the central magnitude, but it
might cause the difference between the isophotal and total magnitudes
to be underestimated at large redshifts. As a further simplification,
I assume all galaxies are face-on and axisymmetric. This assumption
exaggerates the difference between isophotal and total magnitudes.

To determine the isophotal magnitude, $m_{\mathrm{iso}}$, I use a
limiting isophote of about 23 mag/arcsec$^2$, corresponding
approximately to 15\% of the sky brightness in $R$, which are the
stated isophotal limits of \citet{shectman96a}. The total flux within
this isophote is used to calculate the apparent magnitude of the
galaxy in the sample. To determine the central magnitude, I take a
circular aperture of diameter $3.5''$. I add central magnitude errors
with a dispersion of $0.17$, similar to the stated errors in the
photometry of faint galaxies in the survey. I then apply the same flux
and central magnitude cuts on the mock survey as to the real data.

\begin{figure}
\figurenum{\fignum}
\plotone{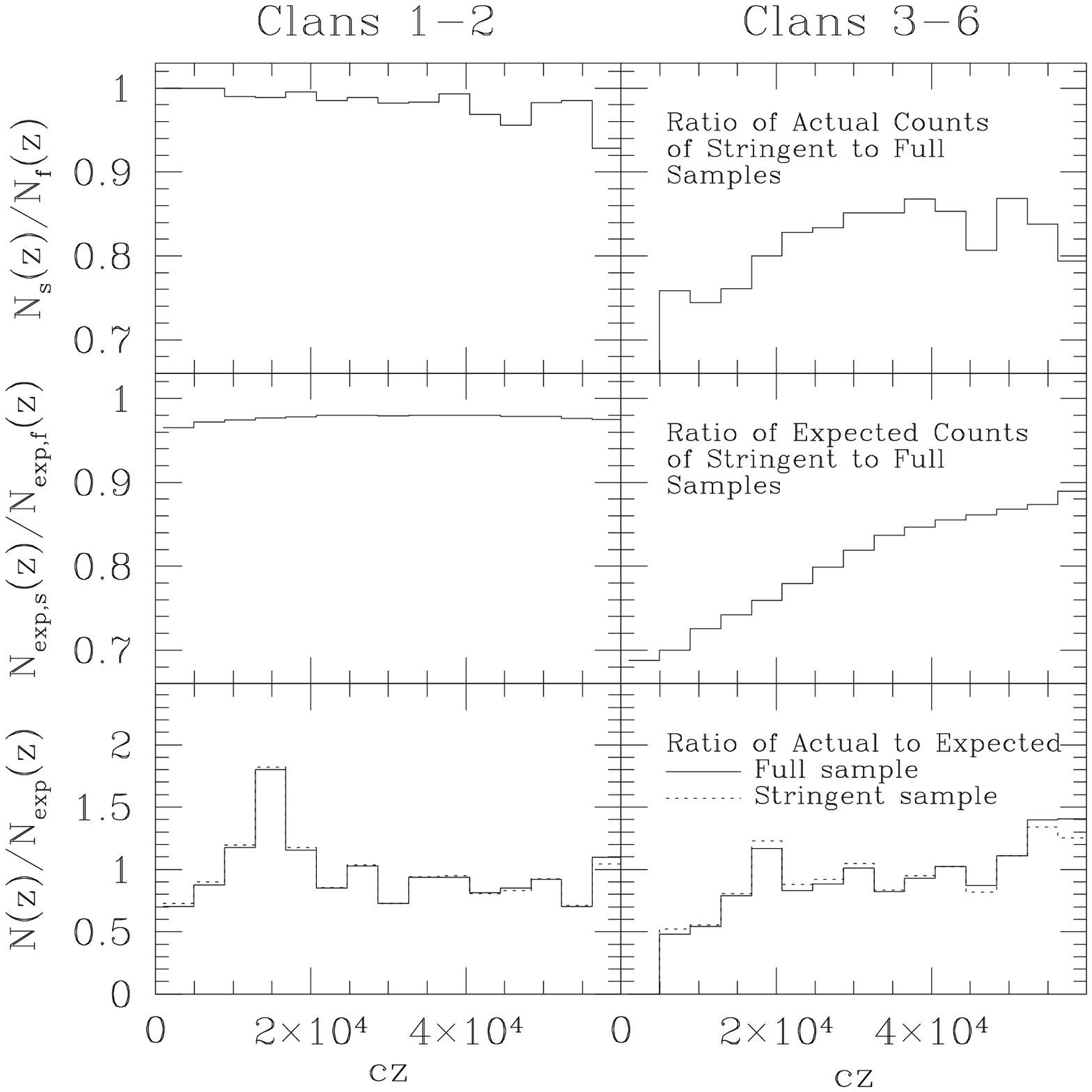}
\caption{\label{c5_deltaz} A comparison of $N(z)$ and
$N_{\mathrm{exp}}(z)$ for the full sample {\it (f)} and the stringent
sample {\it (s)}, for each galaxy type in the N112 fields. The top
panels plot the ratio of the galaxy counts for the stringent sample to
those of the full sample, showing that the surface-brightness cut
preferentially excludes low-redshift, late-type galaxies. The middle
panels plot the ratio of the expected counts of each sample, showing
that the slight decrease in the luminosity function at the faint end
accounts for the dearth of nearby, late-type galaxies. The bottom
panels, which are in some sense implied by the upper two panels, plot
the ratio of actual to expected counts for the full sample {\it (solid
line)} and the stringent sample {\it (dotted line)}. Thus, the derived
density fields of both galaxy types change very little. }
\end{figure}

\begin{figure}
\figurenum{\fignum}
\plotone{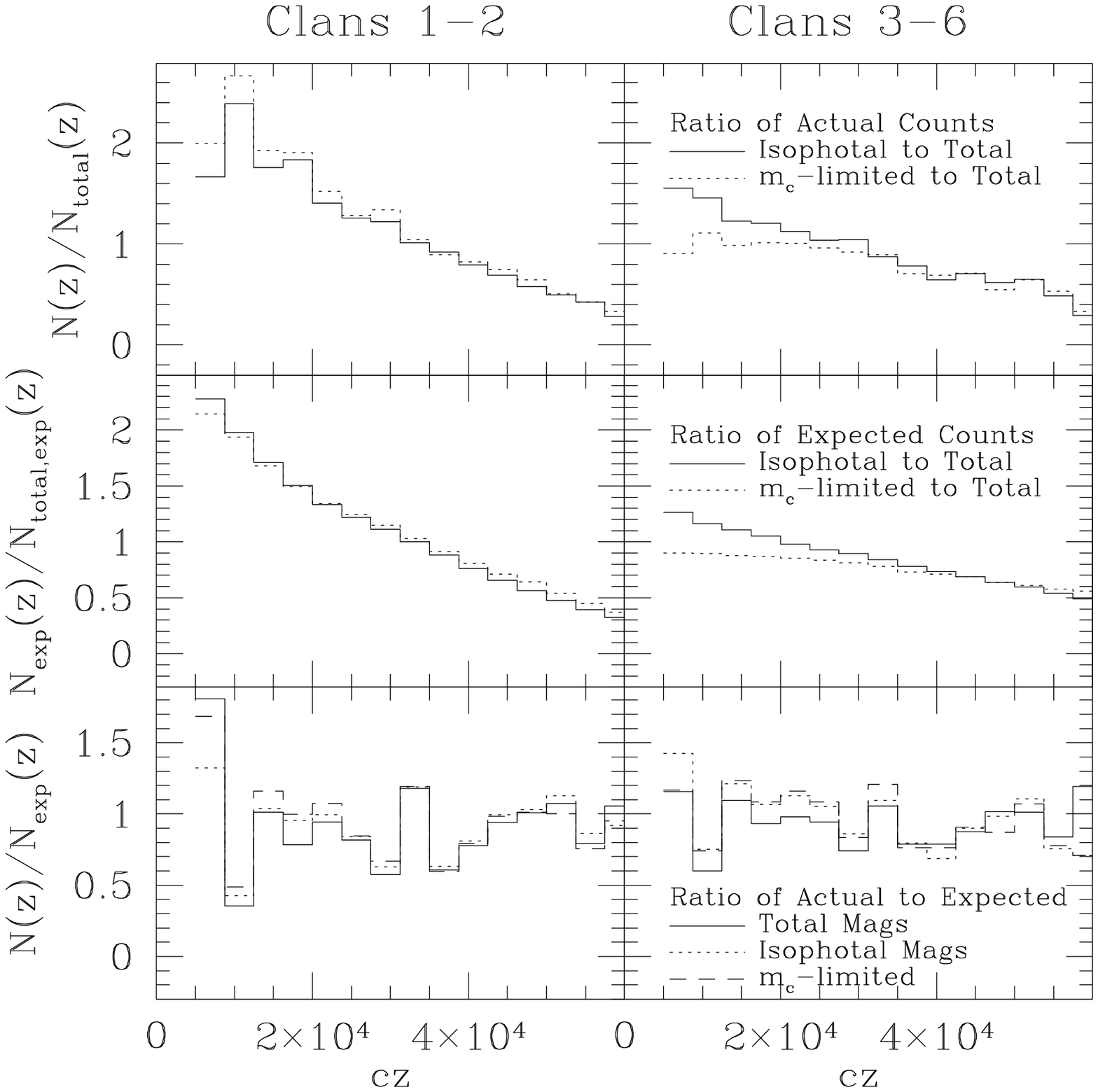}
\caption{\label{c5_mockz} A comparison of $N(z)$ and
$N_{\mathrm{exp}}(z)$ between the sample using total magnitudes, the
sample using isophotal magnitudes, and the sample which uses isophotal
magnitudes {\it and} the $m_c$ cut. Panels have the same meaning as in
Figure \ref{c5_deltaz}. Note that using isophotal magnitudes
eliminates high-redshift, early-type galaxies preferentially, while
placing $m_c$ limits eliminates low-redshift, late-type galaxies
preferentially. However, these elimations change our derived
luminosity function in such a way as to make the selection function
reasonable, so that the density field is not greatly affected.  Note
that since all the samples had nearly the same number of galaxies due
to the finite number of fibers, the ratios in the top two panels are
not constrained to be below unity.}
\end{figure}

I produce three mock catalogs to test these observational effects, all
taken from the same vantage point in the same realization, so I can
compare their density fields directly.  First, I produce a mock
catalog in which the observers have been able to measure the total
magnitudes of the galaxies.  Second, I produce one which is
flux-limited, but based on isophotal magnitudes as described
above. Third, I produce one which uses isophotal magnitudes and is
also $m_c$-limited; that is, the central magnitude cuts have been
applied. I can compare these different catalogs by looking at $N(z)$
as well as $N_{\mathrm{exp}}(z)$ for each, as determined by fitting
for the luminosity function and calculating the selection function; I
make this comparison in Figure \ref{c5_mockz}, which is analogous to
Figure \ref{c5_deltaz} for the observations. Using isophotal
magnitudes evidently causes the systematic elimination of galaxies at
high redshifts, where $(1+z)^4$ dimming and the effects of seeing
start to become important.  Meanwhile, as I showed for the real
observations in Figure \ref{c5_deltaz}, the $m_c$ cut elimates
galaxies at low redshift. However, the changes to the luminosity
function caused by these eliminations seems to cause the selection
functions to be reasonable estimates of the probability of observing a
galaxy at all redshifts. The bottom panels of Figure
\ref{c5_mockz} show, correspondingly, that the density fields of
galaxies are thus unaffected by these changes in galaxy selection.  I
perform the counts-in-cells analysis on these galaxies, and as shown
in Table \ref{c5_iso}, these observational effects do not appear to be
able to cause the sort of stochasticity observed in the real sample.

%
% Conclusions
%

\section{Summmary and Conclusions}
\label{c5_conclusions}

I have presented a straightforward maximum likelihood method to
determine the relationship between the density fields of different
galaxies types on a point-by-point basis by looking at the joint
counts-in-cells distribution $P(N_e, N_l)$. Using mock catalogs, I
have demonstrated the reliability of the method. I have applied the
method to the LCRS in an attempt to constrain the nature of the
segregation of different galaxy spectral types (as classified by
\citealt{bromley98a}). At most a small amount of stochasticity affects
the relationship between early- and late-type galaxies in the LCRS,
corresponding to $r\sim 0.87$, a larger correlation coefficient than
found using the simple moments method of TB99. In addition, it is
likely that even this result is low because of poorly understood
selection effects in the survey, and that the true value of $r$ is
closer to $\sim 0.95$.

In either case, the large scatter predicted by \citet{blanton99a} from
hydrodynamic simuations does not seem to exist, and the results are
more consistent with the semi-analytic predictions of
\citet{somerville99a}. It is not clear yet what the implications of
this result are, but there are at least three possibilities. First,
because the survey is selected in the $R$ band and is
surface-brightness limited, there may not be a sufficient range of
galaxy types represented to reveal the predicted stochasticity. The
fact that the relative bias $b$ between early- and late-type galaxies
is also smaller than predicted (1.2 instead of 1.5) is consistent with
this explanation. Second, since the simulations of \cite{blanton99a}
are low resolution and cannot resolve galactic disks, it may be that
the simulations are not modelling important effects on subgrid scales
which would considerably reduce the stochasticity.  Third and most
interesting (though probably least likely), is the possibility that
the fundamental principles behind the way galaxy formation is
approximated in the simulations are flawed, and need to be
revised. Improved simulations and the analysis of new, larger, and
more complete redshift surveys such as the SDSS will help answer these
questions.

In addition to the main result, I have found suspicious behavior of
the selection function derived for the sample (both my own and that of
\citealt{bromley98a}). A thorough investigation of possible causes of
these errors, using the data itself as well as mock catalogs, has
turned up no likely cause of this effect, including surface-brightness
selection effects, the use of isophotal magnitudes, and errors in the
$K$-correction. On the other hand, it is possible that some of the mock
catalog experiments presented here are misleading because the model I
used for galaxy profiles was inadequate (for instance, if I used an
inaccurate distribution of galaxy sizes).  Analysis of larger surveys
with better quality images and spectra, such as the SDSS, may thus be
more useful than the mock catalogs in understanding the effect. I must
note that the distinct, though unlikely, possibility remains that the
low redshift portion of the LCRS is indeed an unusual section of the
universe, either due to a rapid evolution of galaxy properties between
$z\approx 0.2$ and today or because of some peculiar local phenomenon.

The inadequacy of the selection function may have consequences for
other results based on the LCRS.  First, I have shown here that the
low correlation coefficients measured by TB99 may be in doubt. Second,
the excess large-scale power in this survey claimed by
\citet{landy96a} may be due to this effect. In fact, the largest
amplitude wave in the survey that those authors detect is in the
``outward,'' redshift direction, which might indicate that redshift
dependent selection effects could be contaminating their results; on
the other hand, they also detect large waves tangent to the redshift
direction, which might not be so readily explained.

In any case, the method presented here is applicable to any comparison
of counts-in-cells of different galaxy populations. It may be most
useful in surveys which are volume-limited and have simpler
geometries. Such surveys would also make it easier to explore the
scale-dependence of the relative bias of galaxies; this task is
difficult in the LCRS, since looking at larger scales forces one to
change the geometry of one's cells, which as I have shown affects the
results.  In particular, the Sloan Digital Sky Survey (SDSS;
\citealt{gunn95a}) and the Two-Degree Field (2DF;
\citealt{colless98a}) would allow one to make powerful tests of the
nature of morphological segregation. It is possible, of course, to
compare the galaxy densities in different surveys using this method
(\citealt{seaborne99a}).  For example, one might use the future
$K$-selected redshift survey based on the Two-Micron All Sky Survey
(2MASS; \citealt{beichman98a}) to compare in the appropriate volume to
the SDSS or 2dF.

In conclusion, the details of morphological segregation contain much
information about how galaxies formed. This paper has attempted to
extract some of this information by measuring the stochasticity in the
relative clustering of galaxy types. Future redshift surveys and more
sophisticated galaxy formation models will be able to make much more
powerful and informative tests.

\acknowledgments

Thanks to Michael Strauss for advice on this work as well as comments
on the text. I am indebted to Benjamin Bromley, Marc Davis, Daniel
Koranyi, Huan Lin, Max Tegmark, and Douglas Tucker for extensive
advice and for access to source code.  In addition, I thank Jeremiah
Ostriker, James Gunn, David Spergel, Roman Juszkiewicz, Robert Lupton,
Jim Annis, and Uros Seljak for useful discussions. Michael Way,
Stephane Ethier, and Ryan Scranton did a great job of maintaining the
computer systems on which this work was performed. I am grateful for
the hospitality of the Department of Physics and Astronomy at the
State University of New York at Stony Brook, who kindly provided
computing facilities on my frequent visits there. Finally, this work
would not have been possible without the public availability of the
Las Campanas Redshift Survey data, for which I thank the LCRS
team. This work was supported in part by the grants NAG5-2759,
AST93-18185 and AST96-16901, the Princeton University Research
Board, the DOE, and the NASA grant NAG 5-7092 at Fermilab.

\appendix
\section{Maximum Likelihood Calculation of the Luminosity Function}

The goal of this section is to describe how to maximize the condition
probability:
\begin{equation}
\label{c5_Lzcond2}
p(L_j| z_j) = \frac{p(L_j,z_j)}{p(z_j)} = \frac{\Phi(L_j) f_g(m_j)}
{\int_{L_{{\mathrm{min}},j}}^{L_{{\mathrm{max}},j}} dL \Phi(L) f_g(m)},
\end{equation}
in terms of a model for the luminosity function $\Phi(L)$, using the
method described by \citet{efstathiou88a}.

Any interpolation scheme used to approximate
the factors in Equation (\ref{c5_Lzcond2}) can be expressed as:
\begin{eqnarray}
\Phi(L) &=& \sum_{i=1}^{N_{\mathrm{steps}}} \Phi_i W(L, i),
\mathrm{~and}\cr 
\int_{L_{\mathrm{min}}}^{L_{\mathrm{max}}} dL \Phi(L)
&=& \sum_{i=1}^{N_{\mathrm{steps}}} \Phi_i \left[H(L_{\mathrm{min}},i)
- H(L_{\mathrm{max}},i)\right].
\end{eqnarray}
Here $\Phi_i$ refers to the luminosity function determined in bin $i$,
bounded by $L_i$ and $L_{i+1}$. In the case of piecewise constant
interpolation, which is sufficient for my purposes,
\begin{eqnarray}
W(L,i) &=& \left\{\begin{array}{cl}
1 & {\mathrm{if~}} L_i<L<L_{i+1}, {\mathrm{~and}}\cr
0 & {\mathrm{otherwise}}
\end{array}\right.\cr
H(L,i) &=& \left\{\begin{array}{cc}
0 & {\mathrm{if~}} L_{i+1}<L \cr
L_{i+1}-L & {\mathrm{if~}} L_i<L\le L_{i+1} \cr
L_{i+1}-L_i & {\mathrm{if~}} L<L_{i} \cr 
\end{array}\right.
\end{eqnarray}
The formulae for piecewise linear interpolation are given by
\citet{koranyi97a}. Using these approximations, multiplying the
conditional probabilities given by Equation (\ref{c5_Lzcond2}) for each
galaxy $j$ in the survey, and imposing the maximum likelihood condition
that 
\begin{equation}
\frac{\partial}{\partial \Phi_k} {\mathcal{L}} \equiv
\frac{\partial}{\partial \Phi_k} \sum_{j=1}^{N_{\mathrm{gals}}} \ln
\left[p(L_j| z_j)\right] = 0
\end{equation}
for all of the $N_{\mathrm{step}}$ parameters, yields an iterative
equation for each $\Phi_k$: 
\begin{equation}
\label{c5_lfiter}
\Phi_k = \frac{
\sum_{j=1}^{N_{\mathrm{gals}}} \left[W(L_j, k) \Phi_k f_g(m_k) / 
\sum_{i=1}^{N_{\mathrm{steps}}} \Phi_i f_g(m_i) W(L_j, i)\right]}
{\sum_{j=1}^{N_{\mathrm{gals}}} \left[\left(H(L_{{{\mathrm{min}},j}},k)-
H(L_{{\mathrm{max}},j},k)\right) f_g(m_k) /
\sum_{i=1}^{N_{\mathrm{steps}}} \Phi_i f_g(m_i)
\left(H(L_{{\mathrm{min}},j},i)- H(L_{{\mathrm{max}},j},i)\right)
\right]},
\end{equation}
which reduces to Equation (2.12) of \citet{efstathiou88a} in the case
of piecewise constant interpolation, which I will use here. Note that
for the case of piecewise linear interpolation, this formula differs
from the one given by \citet{koranyi97a}, which is missing terms in the
denominator of the numerator. Experiments have shown that the
difference between using this formula and theirs is fairly small.

As described in \citet{efstathiou88a}, the luminosity function is
derived by starting with some initial guess for the $\Phi_k$ and
iterating until the improvement in the likelihood per iteration is
small. Throughout, one maintains the normalization condition:
\begin{equation}
\label{c5_normeq}
g \equiv \sum_{i=1}^{N_{\mathrm{steps}}} \Phi_i (L_{i+1}-L_i) - 1 = 0.
\end{equation}
The errors in $\ln\Phi_k$ can be  calculated by inverting the matrix:
\begin{equation}
I_{ij} = \left[\begin{array}{cc}
-\frac{\partial^2\ln\mathcal{L}}{\partial\ln\Phi_i\partial\ln\Phi_j}
-\frac{\partial g}{\partial\ln\Phi_i}
\frac{\partial g}{\partial\ln\Phi_j} & 
-\frac{\partial g}{\partial\ln\Phi_i} \cr
-\frac{\partial g}{\partial\ln\Phi_j} & 0
\end{array}\right]
\end{equation}
The diagonal elements of $I_{ij}^{-1}$ are the errors in each
parameter $\Phi_k$, while the off-diagonal elements represent the
covariances.

This procedure has determined the shape of the luminosity function,
but one has yet to determine its amplitude. To do so one must calculate
the selection function, which with the interpolation scheme can be
approximated by
\begin{equation}
\label{c5_sfeq_approx}
\phi(z) = \sum_{i=1}^{N_{\mathrm{steps}}} \Phi_i f_g(m_i) f_{t}
\left(H(L_{\mathrm{min}},i)- H(L_{\mathrm{max}},i)\right).
\end{equation}
The most straightforward estimate of the normalization of the
luminosity function (though not the minimum variance estimator) is
\begin{equation}
n_1 = \frac{1}{V} \sum_{j=1}^{N_{\mathrm{gals}}} \frac{1}{\phi_j},
\end{equation}
where $V$ is the size of the volume probed, and the error can be
estimated as
\begin{equation}
\Avg{\delta n_1^2}^{1/2} = \frac{1}{V} \left[
\sum_{j=1}^{N_{\mathrm{gals}}} \frac{1}{\phi_j^2}\right]^{1/2}
\end{equation}
Given $n_1$, one can express the number of galaxies per unit volume per
unit luminosity by $n_1 \Phi(L)$.

\newpage

\setcounter{thetabs}{0}

%\begin{deluxetable}{ccccr} 
%\tablewidth{0pt}
%\tablecolumns{5}
%%\tablenum{\tabnum}
%\tablecaption{\label{c5_pdffits}
%PDF fits to LCRS counts-in-cells.}
%\tablecomments{The parameter ${\mathcal L}$ is given with respect to
%the Gaussian fit: ${\mathcal L}\equiv- 2 \ln (L/L_{\mathrm{
%gauss}})$. Note that $\sigma_eS_{3,e}\approx\sigma_lS_{3,l}$,
%consistent with linear bias between the two fields.}
%\tablehead{Galaxy Type & PDF Form & $\sigma$ & $S_3$ & ${\mathcal L}$}
%\startdata
%Clans 1--2 & Gaussian & $ 0.59 \pm  0.04$ & --- &   0.0\cr
% & Edgeworth & $ 0.50 \pm  0.02$ & $ 3.03 \pm 0.507$ & $-30.6$ \cr
% & Log-normal & $ 0.54 \pm  0.02$ & --- & $-35.6$ \cr
%Clans 3--6 & Gaussian & $ 0.39 \pm  0.02$ & --- &   0.0\cr
% & Edgeworth & $ 0.39 \pm  0.02$ & $ 3.65 \pm 0.838$ & $-28.5$ \cr
% & Log-normal & $ 0.41 \pm  0.02$ & --- & $-36.9$ \cr
%\enddata
%\end{deluxetable}

\stepcounter{thetabs}
\begin{deluxetable}{cccccrrrr}
%\tabletypesize{\scriptsize}
\rotate
\tablecolumns{9}
\tablenum{\tabnum}
\tablewidth{0pt}
\tablecaption{\label{c5_biasrd} PDF and Bias Fits For Several Ranges
of Redshift}
\tablecomments{The parameter ${\mathcal L}$ is given with respect to
the linear fit: ${\mathcal L}\equiv- 2 \ln (L/L_{\mathrm
{linear}})$. $P_{\mathrm{random}}$ is the probability that a fit this
good is achieved at random, given that the best fit model is correct;
if close to one, the fit is poor, if $\sim 0.5$ or less, the fit is
good. $P_{\mathrm{linear}}$ is the probability of achieving the
observed likelihood ratio with respect to the linear fit assuming that
the linear fit is correct; low values indicate that the given fit is
significantly better than linear.}
\tablenotetext{\ast}{For the second section, I used all the cells, but
determined the expected counts separately for $cz<24,000$ km/s and
$cz>24,000$ km/s, using the luminosity function in each region shown
in Figure \ref{c5_lfhilo}.}
\tablehead{Redshift Range (km/s)& $\sigma_e$ & $\sigma_l$ &
Bias Model & $b_1$ & $b_2$ or $\sigma_b$ & ${\mathcal L}$ 
& $P_{\mathrm{random}}$ & $P_{\mathrm{linear}}$}
\startdata
$10,000 < cz <46,000$ & $ 0.54 \pm  0.02$ & $ 0.41 \pm  0.02$ & 
Linear & $ 0.76 \pm  0.02$ & --- &   0.0 & $>0.995$ & N/A \cr
& & & Power-law & $ 0.75 \pm  0.02$ & --- &   13.0 & $>0.995$ & N/A \cr
& & & Quadratic & $ 0.73 \pm  0.03$ & $ 0.18 \pm  0.03$ & $-13.1$ & 
$>0.995$ & $< 0.005$ \cr
& & & Broken & $ 0.64 \pm 0.05$ & $ 0.89 \pm 0.05$ & $-5.2$ & $>0.995$ &
0.020 \cr
& & & Stochastic & $ 0.63 \pm 0.05$ & $ 0.21 \pm 0.02$ & $-106.0$ &
0.370 & $<0.005$\cr
$10,000 < cz <46,000$\tablenotemark{\ast} 
& $ 0.52 \pm  0.03$ & $ 0.39 \pm  0.02$ & 
Linear & $ 0.77 \pm  0.02$ & --- &   0.0 & $>0.995$ & N/A \cr
& & & Power-law & $ 0.75 \pm  0.03$ & --- &   -4.8 & $>0.995$ & $0.015$ \cr
& & & Quadratic & $ 0.80 \pm  0.03$ & $-0.07 \pm  0.04$ &  -3.7 & $>0.995$ & $0.050$ \cr
& & & Broken & $ 0.89 \pm  0.06$ & $ 0.68 \pm  0.05$ &  -3.8 & $>0.995$ & $0.040$ \cr
& & & Stochastic & $ 0.71 \pm  0.04$ & $ 0.16 \pm  0.02$ & -35.5 & $0.740$ &
$<0.005$ \cr
%$10,000 < cz <36,000$ & $ 0.50 \pm  0.03$ & $ 0.39 \pm  0.02$ & 
%Linear & $ 0.75 \pm  0.03$ & --- &   0.0 & $>0.995$ & N/A \cr
%& & & Power-law & $ 0.72 \pm  0.03$ & --- &   6.2 & $>0.995$ & N/A \cr
%& & & Quadratic & $ 0.73 \pm 0.04$ & $ 0.19 \pm 0.05$ & $-8.2$ & $>0.995$ 
%& 0.005 \cr
%& & & Broken & $ 0.66 \pm 0.05$ & $ 0.85 \pm 0.06$ & $-1.8$ & $>0.995$ &
%0.165 \cr
%& & & Stochastic & $ 0.55 \pm  0.06$ & $ 0.22 \pm  0.02$ & $-119.7$ &
%0.810 & $<0.005$ 
%\cr
%$36,000 < cz <46,000$ & $ 0.61 \pm  0.03$ & $ 0.46 \pm  0.03$ &
%Linear & $ 0.79 \pm  0.05$ & --- &   0.0 & 0.320 & N/A \cr
%& & & Power-law & $ 0.77 \pm  0.05$ & --- &   1.3 & 0.140 & N/A \cr
%& & & Quadratic & $ 0.81 \pm  0.06$ & $-0.06 \pm  0.05$ &  $-0.6$ &
%0.180 & 0.420 \cr
%& & & Broken & $ 0.80 \pm  0.11$ & $ 0.78 \pm  0.08$ &  $-0.0$ & 0.210
%& 0.955 \cr
%& & & Stochastic & $ 0.78 \pm  0.04$ & $ 0.07 \pm  0.05$ &  $-0.2$ &
%0.185 & 0.370 \cr
$24,000 < cz <46,000$ & $ 0.56 \pm  0.03$ & $ 0.43 \pm  0.03$ &
Linear & $ 0.81 \pm  0.03$ & --- &   0.0 & 0.610 & N/A \cr
& & & Power-law & $ 0.80 \pm  0.03$ & --- &   3.1 & 0.480 & N/A \cr
& & & Quadratic & $ 0.81 \pm  0.04$ & $ 0.00 \pm  0.03$ &  $-0.0$
& 0.565 & 0.990 \cr
& & & Broken & $ 0.82 \pm  0.07$ & $ 0.80 \pm  0.05$ &  $-0.0$ & 0.585
& 0.920 \cr
& & & Stochastic & $ 0.77 \pm  0.04$ & $ 0.13 \pm  0.03$ &  $-8.9$ &
0.105 & $<0.005$ \cr 
\enddata  
\end{deluxetable}

\stepcounter{thetabs}
\begin{deluxetable}{cccccrr}
%\tabletypesize{\scriptsize}
\tablewidth{0pt}
\rotate
\tablecolumns{7}
\tablenum{\tabnum}
\tablecaption{\label{c5_benchmark} PDF and Bias Fits for Mock
Catalogs}
\tablecomments{Because of the different cell geometry for the
benchmark sample, the variances are somewhat different. Otherwise, the
results for all the samples, especially for the bias, are remarkably
consistent.}
\tablenotetext{\ast}{
Error bars for the undersampled catalogs are based on the dispersion in
the results for 30 undersampled catalogs, and thus include cosmic
variance as well as the contribution due to statistical errors in the
selection function. They are consistent with the purely statistical
error bars quoted for the other samples.}
\tablehead{Catalog Type & $\sigma_e$ & $\sigma_l$ & Bias Model & $b_1$ 
& $b_2$ or $\sigma_b$ & ${\mathcal L}$}
\startdata
Benchmark &
 $ 0.834 \pm  0.009$ & $ 0.531 \pm  0.005$ & 
Linear & $0.719 \pm 0.004$ & --- &   0.0 \cr
 & & & Quadratic & $0.727 \pm 0.005$ & $-0.009 \pm 0.002$ &  $-6.1$ \cr
 & & & Broken & $0.763 \pm 0.010$ & $0.684 \pm 0.008$ & $-20.8$ \cr
 & & & Stochastic & $0.713 \pm 0.005$ & $0.060 \pm 0.008$ & $-19.4$ \cr
Volume-limited &
 $ 0.63 \pm  0.03$ & $ 0.44 \pm  0.02$ & 
Linear & $ 0.75 \pm  0.02$ & --- &   0.0 \cr
 & & & Quadratic & $ 0.76 \pm  0.02$ & $-0.04 \pm  0.02$ &  $-2.1$ \cr
 & & & Broken & $ 0.82 \pm  0.04$ & $ 0.68 \pm  0.04$ &  $-3.0$ \cr
 & & & Stochastic & $ 0.74 \pm  0.02$ & $ 0.07 \pm  0.04$ &  $-1.3$ \cr
Fully Sampled &
 $ 0.69 \pm  0.03$ & $ 0.48 \pm  0.02$ & 
Linear & $ 0.75 \pm  0.02$ & --- &   0.0 \cr
 & & & Quadratic & $ 0.79 \pm  0.02$ & $-0.07 \pm  0.01$ &  $-9.2$ \cr
 & & & Broken & $ 0.82 \pm  0.05$ & $ 0.69 \pm  0.04$ &  $-3.0$ \cr
 & & & Stochastic & $ 0.74 \pm  0.02$ & $ 0.13 \pm  0.02$ & $-17.5$ \cr
Undersampled\tablenotemark{\ast} &
 $ 0.67 \pm  0.02$ & $ 0.46 \pm  0.02$ & 
Linear & $ 0.75 \pm  0.02$ & --- &   0.0 \cr
 & & & Quadratic & $ 0.76 \pm  0.02$ & $-0.05 \pm  0.04$ &  $-2.4$ \cr
 & & & Broken & $ 0.82 \pm  0.05$ & $ 0.68 \pm  0.06$ &  $-3.3$ \cr
 & & & Stochastic & $ 0.73 \pm  0.02$ & $ 0.08 \pm  0.02$ & $-13.2$ \cr
%Contiguous &  $ 0.69 \pm  0.03$ & $ 0.49 \pm  0.02$ &
%Linear & $ 0.75 \pm  0.02$ & --- &   0.0 \cr
% & & & Quadratic & $ 0.75 \pm  0.02$ & $0.00 \pm  0.01$ &  $-0.0$ \cr
% & & & Broken & $ 0.79 \pm  0.04$ & $ 0.71 \pm  0.03$ &  $-1.6$ \cr
% & & & Stochastic & $ 0.73 \pm  0.02$ & $ 0.08 \pm  0.02$ &  $-5.7$ \cr
%Shallow &  $ 0.61 \pm  0.03$ & $ 0.43 \pm  0.02$ &
%Linear & $ 0.78 \pm  0.02$ & --- &   0.0 \cr
% & & & Quadratic & $ 0.78 \pm  0.02$ & $0.00 \pm  0.02$ &  $-0.0$ \cr
% & & & Broken & $ 0.80 \pm  0.05$ & $ 0.76 \pm  0.03$ &  $-0.3$ \cr
% & & & Stochastic & $ 0.77 \pm  0.03$ & $ 0.07 \pm  0.02$ &  $-3.1$ \cr
\enddata
\end{deluxetable}

\stepcounter{thetabs}
\begin{deluxetable}{cccccrr}
%\tabletypesize{\scriptsize}
\tablewidth{0pt}
\rotate
\tablecolumns{7}
\tablenum{\tabnum}
\tablecaption{\label{c5_iso} PDF and Bias Fits Using Different
Selection Criteria in Mock Catalogs}
\tablecomments{Whether total or isophotal magnitudes are used, and
whether or not the central magnitude limit is applied, seems to have
little effect on the results.}
\tablehead{Catalog Type & $\sigma_e$ & $\sigma_l$ & Bias Model & $b_1$ 
& $b_2$ or $\sigma_b$ & ${\mathcal L}$}
\startdata
Total Magnitudes &
$ 0.70 \pm  0.03$ & $ 0.46 \pm  0.02$ &
Linear & $ 0.73 \pm  0.01$ & --- &   0.0 \cr
 & & & Quadratic & $ 0.74 \pm  0.02$ & $-0.03 \pm  0.01$ &  $-1.3$ \cr
 & & & Broken & $ 0.77 \pm  0.04$ & $ 0.69 \pm  0.03$ &  $-1.1$ \cr
 & & & Stochastic & $ 0.70 \pm  0.02$ & $ 0.11 \pm  0.02$ & $-16.5$ \cr
Isophotal Magnitudes &
$ 0.69 \pm  0.03$ & $ 0.40 \pm  0.02$ &
Linear & $ 0.68 \pm  0.02$ & --- &   0.0 \cr
 & & & Quadratic & $ 0.73 \pm  0.02$ & $-0.13 \pm  0.01$ & $-26.7$ \cr
 & & & Broken & $ 0.84 \pm  0.05$ & $ 0.54 \pm  0.04$ & $-17.5$ \cr
 & & & Stochastic & $ 0.67 \pm  0.02$ & $ 0.09 \pm  0.02$ &  $-9.4$ \cr
$m_c$-limited & 
$ 0.68 \pm  0.03$ & $ 0.44 \pm  0.02$ &
Linear & $ 0.74 \pm  0.02$ & --- &   0.0 \cr
 & & & Quadratic & $ 0.78 \pm  0.02$ & $-0.09 \pm  0.01$ & $-11.4$ \cr
 & & & Broken & $ 0.84 \pm  0.04$ & $ 0.66 \pm  0.04$ &  $-4.6$ \cr
 & & & Stochastic & $ 0.73 \pm  0.02$ & $ 0.10 \pm  0.02$ &  $-8.7$ \cr
\enddata
\end{deluxetable}

%\begin{deluxetable}{ccccccc}
%\tablecolumns{7}
%%\tablenum{\tabnum}
%\tablecaption{\label{c5_config} Properties of PDF and bias fits to
%joint counts-in-cells of early and late type galaxies,
%using different cell configurations in the mock catalogs.}
%\tablehead{Configuration & $\sigma_e$ & $\sigma_l$ & Bias Model & $b_1$ 
%& $b_2$ or $\sigma_b$ & ${\mathcal L}$}
%\startdata
%\enddata
%\end{deluxetable}

\clearpage

\setcounter{thefigs}{0}

%\clearpage
%\begin{figure}
%%\figurenum{\fignum}
%%\plotfiddle{c5_distrib2.ps}{5.5in}{0.}{130.}{130.}{-220in}{-390in}
%\plotone{c5_distrib2.ps}
%\caption{\label{c5_distrib2} Example of how Poisson noise affects
%$P(N_e,N_l)$. The thick grey line shows a deterministic relative bias
%of late-type to early-type galaxies with slope $b_1\approx 0.75$. The
%greyscale indicates the distribution of counts one would expect once
%the relationship $f(\delta_e,\delta_l)$ is convolved with Poisson
%noise, using the expected mean number of galaxies in each cell from the
%LCRS cells described in Section \ref{c5_data}.}
%\end{figure}

\clearpage
\stepcounter{thefigs}

\clearpage

\clearpage
\stepcounter{thefigs}

\clearpage
\stepcounter{thefigs}

\clearpage
\stepcounter{thefigs}

\clearpage
\stepcounter{thefigs}

\clearpage
\stepcounter{thefigs}

%\clearpage
%\begin{figure}
%%\figurenum{\fignum}
%\plotone{c5_lfmccut.ps}
%\caption{\label{c5_lfmccut} Same as Figure \ref{c5_lcrsphi}, for the
%N112 sample alone, and for the full galaxy sample {\it (solid line)}
%and the sample with the more stringent central surface brightness cut
%{\it (dotted line)}. Note that the density of faint, late-type
%galaxies is underestimated in the stringent sample. (The slight
%overestimation of the early-type normalization occurs because most of
%the missing galaxies are late-type, but $f_t$ is calculated for all
%galaxies together, thus causing it to over-correct for missing
%early-type galaxies).}
%\end{figure}

\clearpage
\stepcounter{thefigs}

%\clearpage
%\begin{figure}
%%\figurenum{\fignum}
%\plotone{c5_tilt_ldsplit.ps}
%\caption{\label{c5_tilt_ldsplit} 
%Same as Figure \ref{c5_lcrsldsplit0}, only here the ``expected''
%luminosity distributions are based on luminosity distributions which
%have been ``tilted,'' as described in the text. Note that these severe
%errors result in poorer $\chi^2$ values, but still appear to be a
%decent, if marginal fit, indicating that this statistic is a poor one
%for detecting errors in the luminosity function.}
%\end{figure}

\clearpage
\stepcounter{thefigs}

\clearpage
\stepcounter{thefigs}

\clearpage
\stepcounter{thefigs}

\clearpage
\stepcounter{thefigs}

\clearpage
\stepcounter{thefigs}

%\clearpage
%\begin{figure}
%%\figurenum{\fignum}
%\plotone{c5_catmocke.ps}
%\caption{\label{c5_catmocke} Same as Figure \ref{c5_lcrse}, for the
%early-type galaxies in one of the mock catalogs.}
%\end{figure}

%\clearpage
%\begin{figure}
%%\figurenum{\fignum}
%\plotone{c5_catmockl.ps}
%\caption{\label{c5_catmockl} Same as Figure \ref{c5_lcrse}, for the
%late-type galaxies in one of the mock catalogs.}
%\end{figure}

\clearpage
\stepcounter{thefigs}

\end{document}